\documentclass[runningheads]{llncs}

\usepackage{eccv}

\usepackage{eccvabbrv}

\usepackage{graphicx}
\usepackage{booktabs}
\usepackage{multirow}
\usepackage{pifont}
\usepackage{amsmath,amssymb}

\usepackage[accsupp]{axessibility}  
\usepackage{bbding}

\usepackage{hyperref}

\usepackage{orcidlink}

\begin{document}

\title{Dual-Prior Guided Null-Space Learning with Mixture-of-Splines for Arbitrary Medical Slice Super-Resolution}

\titlerunning{Dual-Prior Null-Space Learning for Arbitrary Slice SR}

\author{
Haofei Song \and
Siyuan Xu \and
Xintian Mao \and
Shaojie Guo \and \\
Qingli Li \and
Yan Wang\textsuperscript{\Envelope}
}

\authorrunning{H.~Song et al.}

\institute{Shanghai Key Laboratory of Multidimensional Information Processing, \\
East China Normal University, Shanghai 200241, China \\
\email{\{hfsong, syxu, 52265904010, 52275904013\}@stu.ecnu.edu.cn, qlli@cs.ecnu.edu.cn, ywang@cee.ecnu.edu.cn}
}
\maketitle

\begin{abstract}
Arbitrary slice super-resolution reconstructs isotropic volumes from anisotropic clinical acquisitions by synthesizing intermediate slices at arbitrary scales.
However, treating this ill-posed inverse problem as unconstrained residual-based regression risks hallucinating anatomically implausible structures or altering the originally observed data.
To address both concerns, this paper presents the \textbf{D}ual-\textbf{P}rior \textbf{N}ull-\textbf{S}pace \textbf{L}earning (DP-NSL) framework, which reformulates the task as a constrained recovery process guided by two complementary priors.
A Measurement-Consistent Projection (MCP) enforces a \emph{Deterministic Observation Prior}: the reconstruction undergoes an exact orthogonal projection that reproduces every acquired slice with zero error, confining all learned details to the unobservable null space.
Within this null space, a Mixture-of-Splines (MoS) module imposes a \emph{Geometric Continuity Prior} by dynamically mixing B-spline experts of different analytic orders, allowing each anatomical region to be modeled with a content-aware level of continuity.
To promote spatial coherence, a Local Spatial Consistency Decoder (LSCD) further injects local inductive bias.
Experiments on three CT and one MRI benchmark show that DP-NSL outperforms existing approaches while strictly preserving measurement consistency. Code is available at \url{https://github.com/DeepMed-Lab-ECNU/Medical-Image-Reconstruction}.
\end{abstract}

\section{Introduction}
\label{sec:intro}

Modern volumetric imaging pipelines, including computed tomography (CT) and magnetic resonance imaging (MRI), routinely produce anisotropic volumes whose through-plane spacing far exceeds the in-plane resolution.
This anisotropy, driven by clinical constraints on acquisition time and radiation dose, compromises downstream tasks such as 3D visualization, medical analysis and diagnosis.
Arbitrary slice super-resolution, the task of synthesizing intermediate slices at arbitrary scales, has therefore become a necessary post-processing step.

\begin{figure}[t]
    \centering
    \includegraphics[width=0.85\textwidth]{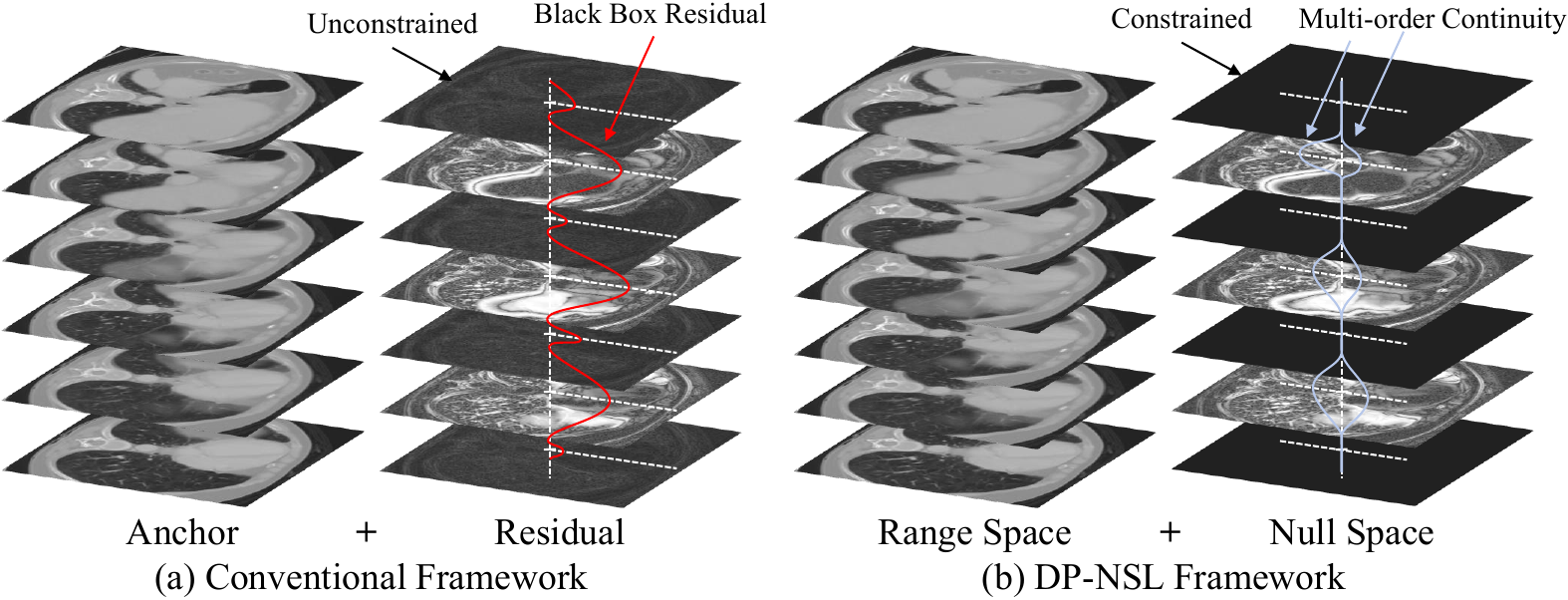}

    \caption{(a) Residual-based methods predict full-resolution outputs without measurement constraints, risking unintended alteration of acquired slices and anatomical hallucinations. (b) The proposed DP-NSL framework structures the reconstruction through two complementary priors: a \emph{Deterministic Observation Prior} that anchors the output to the observed slices, and a \emph{Geometric Continuity Prior} that models unobserved regions with multi-order continuity.}
    \label{fig:motivation}

\end{figure}

Classical interpolation methods (\eg, linear, cubic) offer deterministic stability but fail to recover high-frequency anatomical details, yielding overly smooth reconstructions.
Recent implicit neural representations (INRs)~\cite{chen2021learning} model the volume as a continuous function and achieve superior visual quality. 
However, medical arbitrary slice SR is an ill-posed inverse problem. As shown in \cref{fig:motivation}(a), treating it as unconstrained regression risks hallucinating anatomically implausible structures or altering the originally observed slices. 
Black-box networks alone cannot provide the guarantees needed in clinical settings; instead, the reconstruction should be reframed as a constrained process guided by strong inductive biases. 
The challenge in medical slice SR therefore extends beyond resolution enhancement: the solution must also satisfy \textbf{measurement fidelity} and \textbf{anatomical plausibility}.

To this end, this work presents the \textbf{D}ual-\textbf{P}rior \textbf{N}ull-\textbf{S}pace \textbf{L}earning (DP-NSL) framework, which replaces unconstrained prediction with dual-prior constrained reconstruction: a \emph{Deterministic Observation Prior} that guarantees fidelity to acquired observations, and a \emph{Geometric Continuity Prior} that regularizes the synthesis of unobserved anatomical details.

To enforce the \emph{Deterministic Observation Prior}, the framework introduces a Measurement-Consistent Projection (MCP) grounded in null-range-space decomposition~\cite{wang2023ddnm}.
By projecting the solution onto the observation hyperplane, MCP partitions the reconstruction into a deterministic range-space anchor and an unobserved null-space component. 
This projection mathematically guarantees consistency with the original acquisition: the observed data remain unaltered, while a Null-Space Estimator (NSE) recovers only the missing structures.

Within this null space, synthesizing missing details requires structural regularization to avoid hallucinations.
Biological tissues exhibit heterogeneous smoothness: homogeneous organs such as the liver parenchyma present smooth intensity transitions, whereas structural interfaces like bone-soft tissue boundaries demand sharp representations~\cite{gerig1992nonlinear, birkfellner2014applied}.
To impose the \emph{Geometric Continuity Prior}, this work proposes a Mixture-of-Splines (MoS) module. 
Instead of learning implicit black-box interpolations, MoS explicitly treats continuous B-spline bases~\cite{pak2023btc} as geometric priors.
By dynamically mixing spline experts with different mathematical orders of continuity, the framework adaptively allocates the optimal prior stiffness for each localized anatomical region.

To complement this macroscopic geometric prior, a multi-scale Local Spatial Consistency Decoder (LSCD) is introduced. 
Unlike standard point-wise MLPs that process coordinates independently, LSCD provides a local inductive bias by aggregating neighboring contexts through depthwise convolutions, promoting spatial coherence in the synthesized null-space details.

The main contributions are summarized as follows:
\begin{itemize}
    \item This work formulates arbitrary medical slice SR as a constrained inverse problem and presents the DP-NSL framework, replacing unconstrained resi-dual-based regression with dual-prior guided reconstruction.
    \item By integrating null-range-space decomposition, the proposed MCP provides a hard data-fidelity constraint that keeps every originally observed clinical slice unchanged throughout the reconstruction.
    \item MoS establishes a structural continuity prior within the null space by treating B-splines as explicit geometric priors and adaptively routing smoothness orders according to local anatomy; an LSCD supplies complementary local consistency.
    \item Experiments across CT and MRI datasets show that DP-NSL outperforms existing methods, achieving up to 1.07\,dB PSNR improvement at in-scales while generalizing well to out-of-scales.
\end{itemize}

\section{Related Work}

\subsection{Arbitrary-Scale Image Super-Resolution}

Arbitrary-scale super-resolution (ASSR) aims to reconstruct high-resolution images at any upsampling factor from a single trained model.
Meta-SR~\cite{hu2019meta} first explored this direction by dynamically predicting upsampling convolution weights conditioned on the scale factor.
LIIF~\cite{chen2021learning} then established the dominant paradigm by mapping continuous coordinates and 2D latent features to pixel intensities via a shared MLP.
Subsequent works improve the implicit decoding stage along different axes: LTE~\cite{lee2022local} introduces Fourier-based texture estimation, CiaoSR~\cite{cao2023ciaosr} learns ensemble weights through implicit attention, HIIF~\cite{jiang2025hiif} employs hierarchical positional encoding, and LMF~\cite{he2024lmf} decouples latent modulation from per-pixel rendering for efficiency.
However, LIIF-based methods render each coordinate through a point-wise MLP, limiting their ability to exploit local correlations.
Neural-operator-based approaches~\cite{wei2023srno, luo2024hinote, liu2025difffno} offer an alternative by learning resolution-independent spectral transformations via Fourier neural operators.

A complementary direction estimates parameters of explicit interpolation bases rather than relying on black-box decoders.
BTC~\cite{pak2023btc} predicts B-spline coefficients, knots, and dilations from encoded features, whose compact support and non-negativity naturally suppress the ringing artifacts that affect Fourier-based representations near sharp edges.
BF-STVSR~\cite{kim2025bfstvsr} further validates this principle for temporal interpolation in spatiotemporal video SR.
These results indicate that principled interpolation families, combined with data-driven parameter estimation, can achieve favorable accuracy--efficiency trade-offs, and they motivate the spline-based design adopted here.


\subsection{Medical Volumetric Super-Resolution}
Medical volumetric SR recovers high-resolution 3D images from sub-optimal clinical acquisitions. Distinct from natural images, medical data impose strict anatomical plausibility requirements.
Many existing 3D SR approaches are designed for fixed, integer scale factors. To address the severe anisotropy in routine CT and MRI scans, methods like SAINT~\cite{peng2020saint} and I3Net~\cite{song2024i3net} decompose slice synthesis into specific inter-slice and intra-slice processes. Others~\cite{uhm2025acvtt} propose cross-view texture transfer, using multi-reference non-local attention to improve interpolations.

To enable arbitrary upsampling, ArSSR~\cite{wu2022arbitrary} extends implicit neural representations to medical volumes. Subsequent methods build upon this foundation to improve reconstruction quality. For example, SAINR~\cite{wang2023sainr} enhances feature querying with spatial attention, CycleINR~\cite{fang2024cycleinr} introduces cycle-consistency to reduce over-smoothing, and DC$^2$SR~\cite{zeng2025dc2sr} adopts dual-consistency guidance with curriculum learning.

However, existing INR-based arbitrary-scale methods still rely on coordinate-wise MLPs that overlook local spatial correlations. Without explicit data-fidelity constraints, they also risk altering pixel values at originally observed slices. The proposed framework addresses both limitations through null-space constrained learning and a locally consistent convolutional decoder.

\subsection{Null-Range-Space Decomposition in Inverse Problems}
Null-Range-Space Decomposition (NRSD) provides a mathematically rigorous framework for solving linear inverse problems by separating the solution into a measurement-consistent component (range space) and an unconstrained component (null space).
DDNM~\cite{wang2023ddnm} introduced this idea for zero-shot image restoration, refining null-space details with a pretrained diffusion model while projecting each iterate onto the data-consistency manifold.
NPN~\cite{jacome2025npn} extends the framework with learned nonlinear null-space projections and task-specific regularization, providing convergence guarantees within the plug-and-play paradigm.
Chen~\etal~\cite{chen2025unnull} demonstrate that bilinear interpolation serves as a valid generalized inverse for spectral demosaicing, enabling unsupervised NRSD without ground-truth supervision.

NRSD has been effective in natural image restoration and spectral imaging, but has not yet been applied to arbitrary medical slice SR.
This work adopts NRSD as the theoretical basis of the DP-NSL framework, formulating arbitrary slice SR through a Measurement-Consistent Projection that combines the data-fidelity guarantee of NRSD with a geometry-aware Mixture-of-Splines prior for null-space synthesis.

\section{Method}
\label{sec:method}
\subsection{Problem Formulation and Overall Framework}
\label{sec:method_formulation_overall}
Given an anisotropic 3D volume $\mathbf{V}_{LR}\in\mathbb{R}^{S\times H\times W}$ acquired with coarse through-plane resolution, the goal of arbitrary slice super-resolution is to reconstruct a high-resolution volume ${\mathbf{V}}_{SR}\in\mathbb{R}^{\hat{S}\times H\times W}$ at any upsampling factor $R$, under the supervision of the ground-truth volume $\mathbf{V}_{HR}\in\mathbb{R}^{\hat{S}\times H\times W}$. Let $\mathcal{D}:\mathbb{R}^{\hat{S}\times H\times W}\rightarrow\mathbb{R}^{S\times H\times W}$ denote the downsampling operator modeling the anisotropic slice acquisition. The degradation model is formulated as $\mathbf{V}_{LR} = \mathcal{D}(\mathbf{V}_{HR}) + \boldsymbol{\epsilon}$, where $\boldsymbol{\epsilon}$ represents acquisition noise.

Recovering arbitrary intermediate slices from severely downsampled observations is an ill-posed inverse problem. The unconstrained solution manifold is highly degenerate, and relying on networks to predict target slices without constraints risks hallucinations. The proposed DP-NSL framework therefore formulates the reconstruction as a dual-prior constrained process.

\begin{figure*}[t]
    \centering
    \includegraphics[width=1\textwidth]{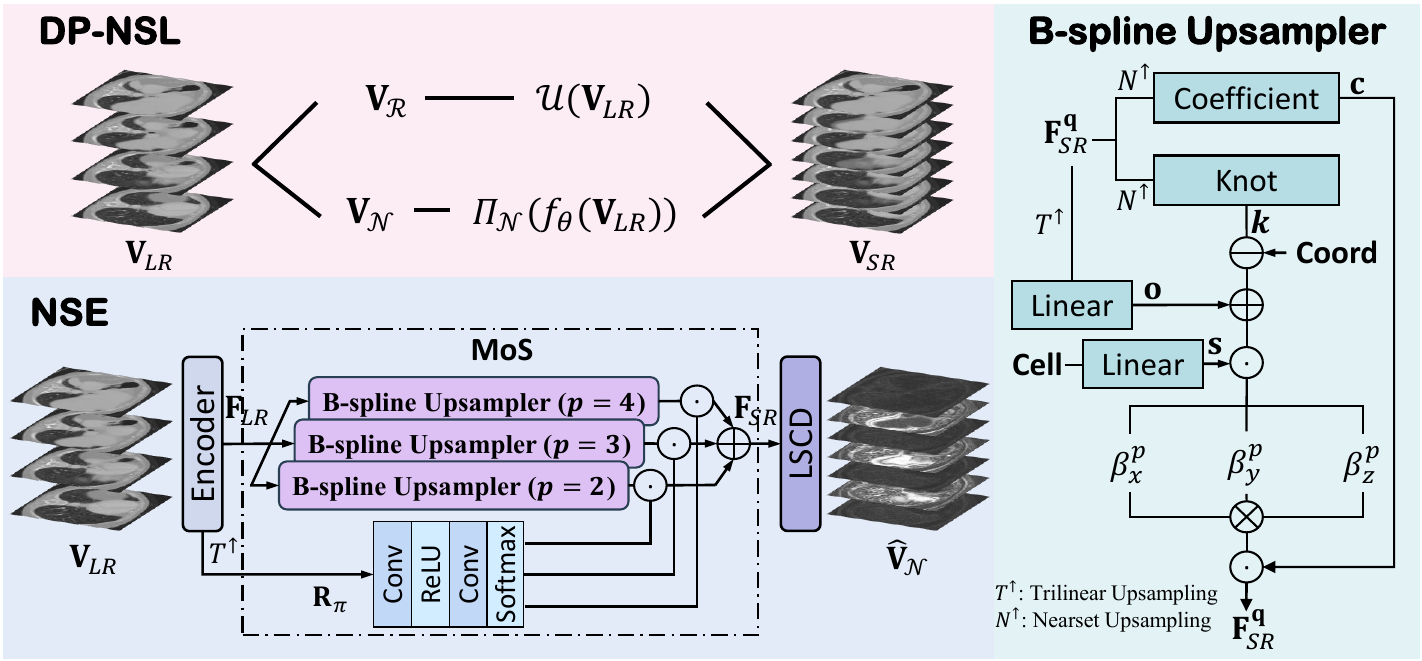}
    \caption{Overview of the proposed Dual-Prior Null-Space Learning (DP-NSL) framework. The reconstruction is decomposed into a deterministic range-space anchor $\mathbf{V}_{\mathcal{R}}$, which preserves observed data, and a learnable null-space component $\mathbf{V}_{\mathcal{N}}$. Within the Null-Space Estimator (NSE), the Mixture-of-Splines (MoS) module enforces spatially adaptive geometric continuity, while a Local Spatial Consistency Decoder (LSCD) injects local inductive bias. 
    }
    \label{fig:arch}

\end{figure*}

As illustrated in \cref{fig:arch}, the forward pipeline is driven by two parallel streams. A deterministic anchor $\mathbf{V}_{\mathcal{R}} = \mathcal{U}(\mathbf{V}_{LR})$ preserves the observed slices, where $\mathcal{U}$ is an upsampling pseudo-inverse. A prior-guided Null-Space Estimator (NSE) $f_\theta$ predicts the unobservable details $\hat{\mathbf{V}}_{\mathcal{N}} = f_\theta(\mathbf{V}_{LR})$. A Measurement-Consistent Projection (MCP) $\Pi_{\mathcal{N}}$ is then applied as a fixed projection, mapping this prediction to the null space to obtain the measurement-safe component $\mathbf{V}_{\mathcal{N}}$. The final super-resolved volume is formed by aggregating these two orthogonal components: ${\mathbf{V}}_{SR} = \mathbf{V}_{\mathcal{R}} + \mathbf{V}_{\mathcal{N}}$.

To regularize the prediction of $\hat{\mathbf{V}}_{\mathcal{N}}$, the NSE incorporates structural priors. It comprises three stages: a 3D encoder to extract spatial features $\mathbf{F}_{LR}$, a Mixture-of-Splines (MoS) module as an upsampler that enforces a continuous geometric prior, and a Local Spatial Consistency Decoder (LSCD) that provides local inductive bias for spatial coherence.

\subsection{Deterministic Observation Prior via Orthogonal Projection}
\label{sec:method_nrsd}
The first principle in medical image reconstruction is \emph{data fidelity}: the super-resolved volume must strictly agree with the clinically acquired slices. However, existing methods~\cite{wu2022arbitrary,wang2023sainr} directly optimize an unconstrained mapping $\mathbb{R}^{S\times H\times W}\rightarrow\mathbb{R}^{\hat{S}\times H\times W}$. Without mathematical constraints, the network is prone to perturbing the pixel values at the originally observed positions.

To enforce the proposed \emph{Deterministic Observation Prior}, this unconstrained process is reformulated into an exact orthogonal projection constrained to the observation subspace via Null-Range-Space Decomposition~\cite{wang2023ddnm}. 
Given the degradation operator $\mathcal{D}$, the constrained reconstruction requires $\mathcal{D}(\mathbf{V}_{SR}) = \mathbf{V}_{LR}$.
Let $\mathcal{U}: \mathbb{R}^{S\times H\times W}\rightarrow\mathbb{R}^{\hat{S}\times H\times W}$ be a pseudo-inverse satisfying the consistency condition $\mathcal{D}\,\mathcal{U} = \mathbf{I}$.
For any target-resolution volume $\mathbf{V}_{SR}$, applying the identity matrix yields an exact orthogonal decomposition:
\begin{equation}
\label{eq:nrsd}
\mathbf{V}_{SR} = \mathcal{U}\mathcal{D}(\mathbf{V}_{SR}) + (\mathbf{I} - \mathcal{U}\mathcal{D})\mathbf{V}_{SR} = \mathcal{U}(\mathbf{V}_{LR}) + (\mathbf{I} - \mathcal{U}\mathcal{D})\mathbf{V}_{SR}.
\end{equation}
In this formulation, the first term $\mathcal{U}(\mathbf{V}_{LR})$ is deterministically measurable from the low-resolution input. Conversely, the second term resides in the null space of the degradation operator and involves the unobserved ideal full-resolution volume.

To resolve this missing information, the unobserved volume $\mathbf{V}_{SR}$ in the null-space term is approximated by the structural output of the NSE, denoted as $\hat{\mathbf{V}}_{\mathcal{N}} = f_\theta(\mathbf{V}_{LR})$. Letting the Measurement-Consistent Projection (MCP) operator be defined as $\Pi_{\mathcal{N}} \triangleq \mathbf{I} - \mathcal{U}\mathcal{D}$, a valid constrained solution manifold is established:
\begin{equation}
\label{eq:constrained_solution}
\mathbf{V}_{SR} = \underbrace{\mathcal{U}(\mathbf{V}_{LR})}_{\mathbf{V}_{\mathcal{R}}:\text{ deterministic anchor}}\; +\; \underbrace{\Pi_{\mathcal{N}}\big(\hat{\mathbf{V}}_{\mathcal{N}}\big)}_{\mathbf{V}_{\mathcal{N}}:\text{ null-space component}}.
\end{equation}

Because $\mathcal{D}\,\Pi_{\mathcal{N}} = \mathcal{D} - \mathcal{D}\mathcal{U}\mathcal{D} = \mathcal{D} - \mathcal{D} = \mathbf{0}$, any signal in the range of $\Pi_{\mathcal{N}}$ is mathematically invisible to the forward operator $\mathcal{D}$.
The range-space component $\mathbf{V}_{\mathcal{R}}$ is computed from observations and is the immutable anchor.
The null-space component $\mathbf{V}_{\mathcal{N}} = (\mathbf{I}-\mathcal{U}\mathcal{D})\hat{\mathbf{V}}_{\mathcal{N}}$ is confined to the null subspace of $\mathcal{D}$, within which the network synthesizes unobserved anatomical details. 

This derivation provides a hard data-fidelity constraint. Regardless of how the network predicts $\hat{\mathbf{V}}_{\mathcal{N}}$, MCP confines all synthetic details to unobserved regions.

\subsection{Prior-Guided Null-Space Refinement}
\label{sec:method_geometry_learning}
While MCP locks the reconstruction to observed data, the unconstrained raw prediction $\hat{\mathbf{V}}_{\mathcal{N}}$ remains vulnerable to anatomical implausibility if learned blindly. To prevent this, the mapping from extracted features $\mathbf{F}_{LR}$ to the high-frequency unobserved details $\hat{\mathbf{V}}_{\mathcal{N}}$ is regularized by two complementary biases: a macroscopic geometric continuity prior and a microscopic local consistency prior.

\subsubsection{Spatially Adaptive Geometric Prior via MoS}
\label{sec:method_moe_bspline}
Medical volumes exhibit heterogeneous smoothness, requiring high-order continuity for homogeneous soft tissues and sharp transitions at anatomical boundaries. Classical INR-based methods map coordinate domains via shared MLPs, imposing a uniform global prior that compromises either smooth regions or sharp edges. To address this, the proposed Mixture-of-Splines replaces unconstrained MLP regression with spatially adaptive geometric interpolation to assign coordinate-specific analytic continuity.

To model arbitrary-scale slice SR, let $\mathbf{F}_{LR} \in \mathbb{R}^{C \times S \times H \times W}$ denote low-reso-lution feature from a 3D encoder and  $\mathbf{q} \in \mathbb{R}^3$ denote a target query coordinate (Coord in Fig.~\ref{fig:arch}) in the high-resolution space. The target high-resolution feature mapping is defined as $\mathbf{F}_{SR}^{\mathbf{q}} = \mathrm{MoS}(\mathbf{F}_{LR}, \mathbf{q})$.

To enforce structural consistency in the continuous spatial field, spatially adaptive B-spline bases are introduced as geometric priors rather than generic interpolators. A B-spline of order $p$ inherently guarantees $C^{p-1}$ analytic continuity~\cite{wang2021spe}. For a specific query $\mathbf{q}$, rather than relying on an MLP, the continuous mapping is evaluated by a geometric spline expert $\mathcal{U}_{p}$. The upsampled feature evaluated at $\mathbf{q}$ is formulated as a separable tensor product:
\begin{equation}
\label{eq:bspline_basis}
    \mathcal{U}_{p}(\mathbf{F}_{LR};\mathbf{q}) = \mathbf{c} \odot \big( \mathbf{b}_x \otimes \mathbf{b}_y \otimes \mathbf{b}_z \big).
\end{equation}
Here, $\otimes$ denotes the standard tensor product constructing the $M^3$-dimensional separable 3D basis, and $\odot$ is element-wise multiplication. The 1D B-spline basis vector along spatial axis $d \in \{x,y,z\}$ is computed as $\mathbf{b}_d = \beta^p\big( (\Delta q_d - \mathbf{k}_d) \odot \mathbf{s}_d \big)$. The relative projection distance $\Delta \mathbf{q} = \mathbf{q} - \mathbf{q}_{LR} + \mathbf{o}$ incorporates a learned sub-voxel offset $\mathbf{o} \in \mathbb{R}^3$ for grid misalignment compensation, where $\mathbf{q}_{LR}$ is the nearest counterpart in the LR grid. The geometric variables are projected from the immediate feature neighborhood surrounding $\mathbf{q}_{LR}$ through shallow convolution layers: structurally-aware scaling coefficients $\mathbf{c} \in \mathbb{R}^{M^3}$, adaptive knots $\mathbf{k}_d \in \mathbb{R}^M$, and continuous dilations $\mathbf{s}_d \in \mathbb{R}^M$ projected from a step size (Cell in Fig.~\ref{fig:arch}) of the HR grid.

Instead of rigidly committing to a single uniform continuous order, MoS dynamically allocates spatially adaptive prior stiffness. A bank of $K$ spline experts $\{\mathcal{U}_{p_k}\}_{k=1}^K$ is instantiated with distinct analytic orders $p_k$. Simultaneously, a lightweight routing network calculates the mixing weights $\boldsymbol{\pi}(\mathbf{q}) \in \mathbb{R}^K$ from the locally interpolated features $\tilde{\mathbf{F}}_{LR}^{\mathbf{q}}$:
\begin{equation}
\label{eq:routing}
    \boldsymbol{\pi}(\mathbf{q}) = \mathrm{Softmax}\big(\mathbf{R}_{\pi}(\tilde{\mathbf{F}}_{LR}^{\mathbf{q}})\big).
\end{equation}
The final high-resolution feature at $\mathbf{q}$ is a weighted mixture of the geometric priors:
\begin{equation}
\label{eq:mos_aggregation}
\mathbf{F}_{SR}^{\mathbf{q}}=\sum_{k=1}^{K} \pi_k(\mathbf{q})\,\mathcal{U}_{p_k}\big(\mathbf{F}_{LR};\mathbf{q}\big).
\end{equation}
MoS thus produces a content-aware continuous spatial representation with locally adapted geometric priors.

\subsubsection{Local Consistency Prior via LSCD}
\label{sec:method_conv_decoder}
By evaluating $\mathbf{F}_{SR}^{\mathbf{q}}$ over the target high-resolution coordinate, a high-resolution dense feature volume $\mathbf{F}_{SR} \in \mathbb{R}^{C \times \hat{S} \times H \times W}$ is finally formed. 
Typical point-wise MLPs process each coordinate independently when mapping this spatial field to the raw unobserved details $\hat{\mathbf{V}}_{\mathcal{N}}$, which can induce micro-level local inconsistencies. To address this and supplement the macroscopic MoS prior, a multi-scale Local Spatial Consistency Decoder is introduced to inject a local inductive bias.

Based on an Inception-style split-transform-merge mechanism~\cite{szegedy2015going,szegedy2016rethinking}, $\mathbf{F}_{SR}$ is partitioned along the channel dimension into parallel sub-groups: $\mathbf{f}_{\text{id}}, \mathbf{f}_{\text{conv}}, \mathbf{f}_1, \mathbf{f}_2$, and $ \mathbf{f}_3$.
The identity group $\mathbf{f}_{\text{id}}$ skips computation to preserve the high-frequency base signal. The $\mathbf{f}_{\text{conv}}$ branch operates as a standard 3D convolution, while the remaining sibling groups are processed through depthwise 3D convolutions with monotonically increasing kernel sizes to capture multi-scale spatial context:
\begin{equation}
\label{eq:lsc_decoder}
    \mathbf{F}_{\text{out}} = \big[ \mathbf{f}_{\text{id}},\; \mathcal{W}_{3}(\mathbf{f}_{\text{conv}}),\; \mathcal{W}_{3}^{\text{DW}}(\mathbf{f}_1),\; \mathcal{W}_{5}^{\text{DW}}(\mathbf{f}_2),\; \mathcal{W}_{7}^{\text{DW}}(\mathbf{f}_3) \big],
\end{equation}
where $[\cdot]$ denotes concatenation along the channel dimension, $\mathcal{W}_{k}$ represents a $k \times k \times k$ standard convolution, and $\mathcal{W}_{k}^{\text{DW}}$ denotes a depthwise spatial convolution. 
After cascading multiple such blocks, a final projection collapses the refined multi-scale feature maps to the raw unobservable signal $\hat{\mathbf{V}}_{\mathcal{N}}$, subsequently bounded by the hard MCP constraint.

\begin{table}[t]
\centering
\caption{Quantitative comparison at in-scales ($\times$2, $\times$3, $\times$4) on four datasets. Best results in \textbf{bold}, second best \underline{underlined}.}
\label{tab:result_inscale}

\renewcommand{\arraystretch}{1}

\resizebox{1\linewidth}{!}{
\begin{tabular}{l|ccc|ccc}
\hline
\multirow{2}{*}{Method} & \multicolumn{3}{c|}{Colon} & \multicolumn{3}{c}{Liver} \\
\cline{2-7}
 & $\times 2$ & $\times 3$ & $\times 4$ & $\times 2$ & $\times 3$ & $\times 4$ \\
\hline
EDSR3D   & \underline{41.84}/{0.9809} & 37.71/0.9600 & \underline{35.37}/\underline{0.9421} & 40.62/0.9733 & 37.02/0.9507 & \underline{34.86}/\underline{0.9307} \\
MetaSR   & 41.26/0.9790 & 37.22/0.9567 & 34.81/0.9361 & 40.76/0.9734 & 36.61/0.9461 & 34.25/0.9220 \\
LTE      & 41.72/0.9806 & 37.62/0.9597 & 35.15/0.9399 & 41.40/0.9766 & \underline{37.15}/\underline{0.9512} & 34.71/0.9281 \\
HIIF     & 41.83/\underline{0.9810} & {37.75}/\underline{0.9604} & 35.26/0.9411 & \underline{41.42}/\underline{0.9768} & \underline{37.15}/0.9512 & 34.69/0.9280 \\
ArSSR    & 40.67/0.9771 & 36.79/0.9534 & 34.24/0.9293 & 40.90/0.9737 & 36.69/0.9452 & 34.18/0.9190 \\
SAINR    & 41.48/0.9798 & \underline{37.78}/0.9603 & 35.30/0.9408 & 41.32/0.9763 & 37.14/0.9509 & 34.68/0.9273 \\
CycleINR & 41.35/0.9794 & 37.38/0.9574 & 34.87/0.9360 & 40.88/0.9732 & 36.70/0.9445 & 34.33/0.9194 \\
DC2SR    & 41.29/0.9794 & 37.40/0.9580 & 34.94/0.9374 & 41.18/0.9763 & 36.96/0.9501 & 34.50/0.9261 \\
\textbf{DP-NSL} & \textbf{42.55/0.9832} & \textbf{38.43/0.9648} & \textbf{35.95/0.9472} & \textbf{42.49/0.9801} & \textbf{38.06/0.9572} & \textbf{35.47/0.9351} \\
\hline
\multirow{2}{*}{Method} & \multicolumn{3}{c|}{Hepatic Vessels} & \multicolumn{3}{c}{IXI} \\
\cline{2-7}
 & $\times 2$ & $\times 3$ & $\times 4$ & $\times 2$ & $\times 3$ & $\times 4$ \\
\hline
EDSR3D   & \underline{43.25}/\underline{0.9847} & \underline{38.97}/\underline{0.9671} & \underline{36.71}/\underline{0.9520} & 47.08/0.9902 & 43.45/0.9788 & \underline{41.47}/\underline{0.9679} \\
MetaSR   & 42.04/0.9813 & 37.87/0.9605 & 35.63/0.9428 & 46.67/0.9893 & 42.87/0.9759 & 40.75/0.9625 \\
LTE      & 42.64/0.9832 & 38.43/0.9640 & 36.13/0.9472 & 47.11/\underline{0.9903} & 43.40/0.9786 & 41.33/0.9669 \\
HIIF     & 43.17/0.9846 & 38.93/0.9670 & 36.68/0.9518 & \underline{47.12}/\underline{0.9903} & 43.44/0.9787 & 41.38/0.9672 \\
ArSSR    & 41.89/0.9812 & 37.85/0.9605 & 35.57/0.9420 & 46.51/0.9888 & 42.70/0.9748 & 40.45/0.9600 \\
SAINR    & 42.68/0.9833 & 38.79/0.9661 & 36.52/0.9504 & 46.47/0.9887 & 43.20/0.9775 & 41.20/0.9658 \\
CycleINR & 42.33/0.9826 & 38.33/0.9636 & 36.05/0.9464 & 46.75/0.9895 & 43.16/0.9773 & 41.04/0.9646 \\
DC2SR    & 42.64/0.9832 & 38.54/0.9646 & 36.23/0.9478 & 47.09/0.9902 & \underline{43.52}/\underline{0.9791} & 41.41/0.9675 \\
\textbf{DP-NSL} & \textbf{43.40/0.9854} & \textbf{39.23/0.9690} & \textbf{36.93/0.9543} & \textbf{47.54/0.9912} & \textbf{43.93/0.9810} & \textbf{41.92/0.9710} \\
\hline
\end{tabular}
}
\end{table}

\section{Experiments}
\label{sec:experiments}

\subsection{Experimental Setup}
\label{sec:exp_setup}

\noindent\textbf{Datasets.}
Evaluation is conducted on both CT and MRI modalities to test cross-modality generalization.
For CT, 3D volumes are collected from the Medical Segmentation Decathlon (MSD)~\cite{antonelli2022medical}: Colon (190 volumes), Liver (201 volumes), and Hepatic Vessels (433 volumes), all exhibiting clinically anisotropic spacing.
For MRI, 185 T1-weighted brain scans acquired at Hammersmith Hospital are selected from the IXI dataset~\cite{ixi_dataset}.
Each dataset is split into training and test subsets following \cite{peng2020saint,song2024i3net}.
Low-resolution inputs are synthesized by directly downsampling the high-resolution volume by a factor $R$ along the slice axis while preserving in-plane resolution.

\noindent\textbf{Implementation Details.}
\label{sec:impl_details}
The encoder adopts a 3D EDSR~\cite{lim2017enhanced}.
The MoS module instantiates three B-spline experts (orders $p\in\{2,3,4\}$).
All models are optimized with Adam~\cite{kingma2014adam} ($\beta_1=0.9$, $\beta_2=0.999$) and an $\ell_1$ reconstruction loss on NVIDIA RTX 3090 GPUs.
The initial learning rate is $1\times10^{-4}$, halved every 200 epochs over a total of 1\,000 epochs, with a batch size of 8.
During training, each volume is center-cropped to $256\times256$ in the axial plane, and a training sample consists of four consecutive LR slices randomly extracted along the slice axis.

\noindent\textbf{Compared Methods.}
The proposed method is compared against approaches from three categories:
(i) {fixed-scale volumetric SR}: EDSR3D~\cite{lim2017enhanced}, which trains a specific model for each factor;
(ii) {generic ASSR}: MetaSR~\cite{hu2019meta}, LTE~\cite{lee2022local}, and HIIF~\cite{jiang2025hiif};
and (iii) {medical ASSR}: ArSSR~\cite{wu2022arbitrary}, SAINR~\cite{wang2023sainr}, CycleINR~\cite{fang2024cycleinr}, and DC$^2$SR~\cite{zeng2025dc2sr}.
All arbitrary-scale methods adopt the same encoder EDSR3D and are jointly trained on $R\in\{2,3,4\}$.
Evaluation covers {in-scales} ($R\in\{2,3,4\}$) and {out-of-scales} ($R\in\{5,6,7\}$), which tests generalization to factors never encountered during training.
The reconstruction performance is evaluated using the Peak Signal-to-Noise Ratio (PSNR) and the Structural Similarity Index (SSIM).

\subsection{Comparison with State-of-the-Art}
\label{sec:comparison}

\noindent\textbf{In-Scales.}
Table~\ref{tab:result_inscale} reports PSNR and SSIM at in-scales  across all datasets.
The proposed DP-NSL consistently achieves the highest performance.
On Colon, improvements over the second-best method reach 0.71~dB at $\times$2 and 0.65~dB at $\times$3;
on Liver, the margin widens to 1.07~dB at $\times$2.
Hepatic Vessels dataset benefits from the spatially adaptive MoS prior, yielding 0.26~dB at $\times$3.
On the IXI brain MRI benchmark, a gain of 0.42~dB at $\times$2 and 0.45~dB at $\times$4 shows that the dual-prior formulation transfers across modalities.
EDSR3D, trained separately at each scale, remains competitive at individual factors but cannot generalize to other scales.
INR-based medical methods such as SAINR and DC$^2$SR perform favorably but lack explicit data-fidelity constraints, leaving observed slices vulnerable to subtle alteration.

\begin{table}[t]
\centering
\caption{Quantitative comparison at out-of-scales ($\times$5, $\times$6, $\times$7) on four datasets. All methods are trained only on $\times$2, $\times$3, $\times$4. Best in \textbf{bold}, second best \underline{underlined}.}
\label{tab:result_outscale}

\renewcommand{\arraystretch}{1}

\resizebox{1\linewidth}{!}{
\begin{tabular}{l|ccc|ccc}
\hline
\multirow{2}{*}{Method} & \multicolumn{3}{c|}{Colon} & \multicolumn{3}{c}{Liver} \\
\cline{2-7}
 & $\times 5$ & $\times 6$ & $\times 7$ & $\times 5$ & $\times 6$ & $\times 7$ \\
\hline
MetaSR   & 33.10/0.9145 & 32.07/0.9012 & 31.11/0.8879 & 32.71/0.9006 & 31.68/0.8821 & 31.02/0.8723 \\
LTE      & 33.44/0.9204 & \underline{32.33}/0.9066 & \underline{31.34}/\underline{0.8931} & \underline{33.11}/0.9075 & \underline{32.13}/\underline{0.8928} & \underline{31.39}/\underline{0.8812} \\
HIIF     & 31.59/0.8537 & 30.72/0.8461 & 30.89/0.8716 & 31.14/0.8202 & 30.57/0.8213 & 30.85/0.8564 \\
ArSSR    & 32.78/0.9114 & 31.66/0.8959 & 30.74/0.8830 & 32.70/0.8993 & 31.61/0.8831 & 30.92/0.8717 \\
SAINR    & \underline{33.55}/\underline{0.9223} & \underline{32.33}/\underline{0.9070} & 31.27/0.8929 & 33.07/\underline{0.9077} & 31.96/0.8923 & 31.22/0.8810 \\
CycleINR & 32.88/0.9119 & 31.80/0.8972 & 30.87/0.8847 & 32.72/0.8976 & 31.53/0.8824 & 30.92/0.8718 \\
DC2SR    & 33.16/0.9168 & 31.91/0.9001 & 30.93/0.8870 & 32.88/0.9051 & 31.90/0.8888 & 31.16/0.8772 \\
\textbf{DP-NSL} & \textbf{34.05/0.9286} & \textbf{32.80/0.9141} & \textbf{31.76/0.9008} & \textbf{33.72/0.9167} & \textbf{32.55/0.9011} & \textbf{31.68/0.8885} \\
\hline
\multirow{2}{*}{Method} & \multicolumn{3}{c|}{Hepatic Vessels} & \multicolumn{3}{c}{IXI} \\
\cline{2-7}
 & $\times 5$ & $\times 6$ & $\times 7$ & $\times 5$ & $\times 6$ & $\times 7$ \\
\hline
MetaSR   & 33.97/0.9240 & 32.95/0.9121 & 32.17/0.9016 & 38.97/0.9461 & 37.75/0.9318 & 36.84/0.9192 \\
LTE      & 34.44/0.9303 & 33.28/0.9178 & 32.47/0.9070 & 39.52/0.9516 & \underline{38.27}/0.9383 & \underline{37.32}/0.9262 \\
HIIF     & 32.12/0.8522 & 31.27/0.8408 & 32.04/0.8809 & 38.91/0.9448 & 37.53/0.9241 & 37.05/0.9197 \\
ArSSR    & 33.98/0.9257 & 32.79/0.9115 & 32.03/0.9014 & 38.81/0.9446 & 37.51/0.9291 & 36.61/0.9166 \\
SAINR    & \underline{34.76}/\underline{0.9345} & \underline{33.46}/\underline{0.9203} & \underline{32.58}/\underline{0.9095} & \underline{39.55}/\underline{0.9525} & 38.22/\underline{0.9387} & 37.26/\underline{0.9268} \\
CycleINR & 33.93/0.9242 & 32.81/0.9107 & 32.10/0.9019 & 38.98/0.9460 & 37.75/0.9321 & 36.89/0.9204 \\
DC2SR    & 34.36/0.9299 & 32.97/0.9139 & 32.20/0.9044 & 39.53/0.9524 & 38.10/0.9372 & 37.19/0.9258 \\
\textbf{DP-NSL} & \textbf{35.20/0.9393} & \textbf{33.98/0.9260} & \textbf{33.04/0.9146} & \textbf{40.17/0.9578} & \textbf{38.80/0.9442} & \textbf{37.69/0.9303} \\
\hline
\end{tabular}
}
\end{table}

\begin{table}[t]
\centering
\caption{Results ($\times 5$) on the RPLHR-CT real dataset.}

\label{tab:realdata}
\renewcommand{\arraystretch}{1}
\setlength{\tabcolsep}{3pt}
\resizebox{1\linewidth}{!}{
\begin{tabular}{lccccccccc}
\toprule
Metric & EDSR3D & MetaSR & LTE & HIIF & ArSSR & SAINR & CycleINR & DC2SR & \textbf{DP-NSL} \\
\midrule
PSNR & 29.63 & 30.92 & \underline{31.20} & 21.83 & 28.06 & 30.66 & 29.07 & 27.44 & \textbf{34.91} \\
SSIM & 0.8342 & 0.8444 & \underline{0.8505} & 0.7187 & 0.8093 & 0.8384 & 0.8205 & 0.7926 & \textbf{0.8925} \\
\bottomrule
\end{tabular}
}

\end{table}

\noindent\textbf{Out-of-Scales.}
Table~\ref{tab:result_outscale} evaluates generalization at unseen factors $\times$5, $\times$6, and $\times$7.
DP-NSL maintains a consistent lead over all baselines across all datasets.
On Hepatic Vessels, improvements exceed 0.44~dB at every unseen factor, with the gap widening at the most challenging $\times$7.
On IXI, the margin is up to 0.62~dB at $\times$5.
HIIF, despite strong in-scales results, suffers severe degradation at unseen scales.
In contrast, DP-NSL shows a stable geometric prior: the inherent continuity of B-spline bases holds regardless of sampling density, allowing principled extrapolation to unseen upsampling factors.

\noindent\textbf{Real-Data Validation.}
Further experiments on the paired real dataset RPLHR-CT~\cite{yu2022rplhr} test practical validity of the method. 
As shown in Table~\ref{tab:realdata}, DP-NSL achieves 34.91 dB PSNR at $\times$5.

\begin{figure}[t]
    \centering
    \includegraphics[width=1\textwidth]{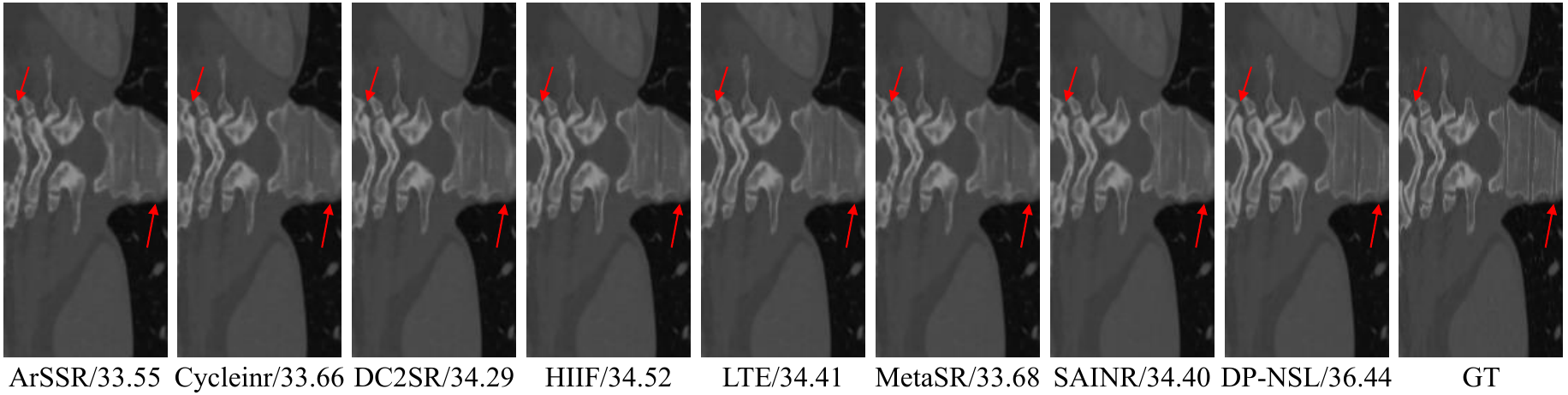}
    \caption{Visualization on the Liver dataset (sagittal view, $\times$3). The sagittal plane directly reveals through-plane reconstruction quality. 
    }
    \label{fig:visual_liver_sag}

\end{figure}

\begin{figure*}[t]
    \centering
    \includegraphics[width=1\textwidth]{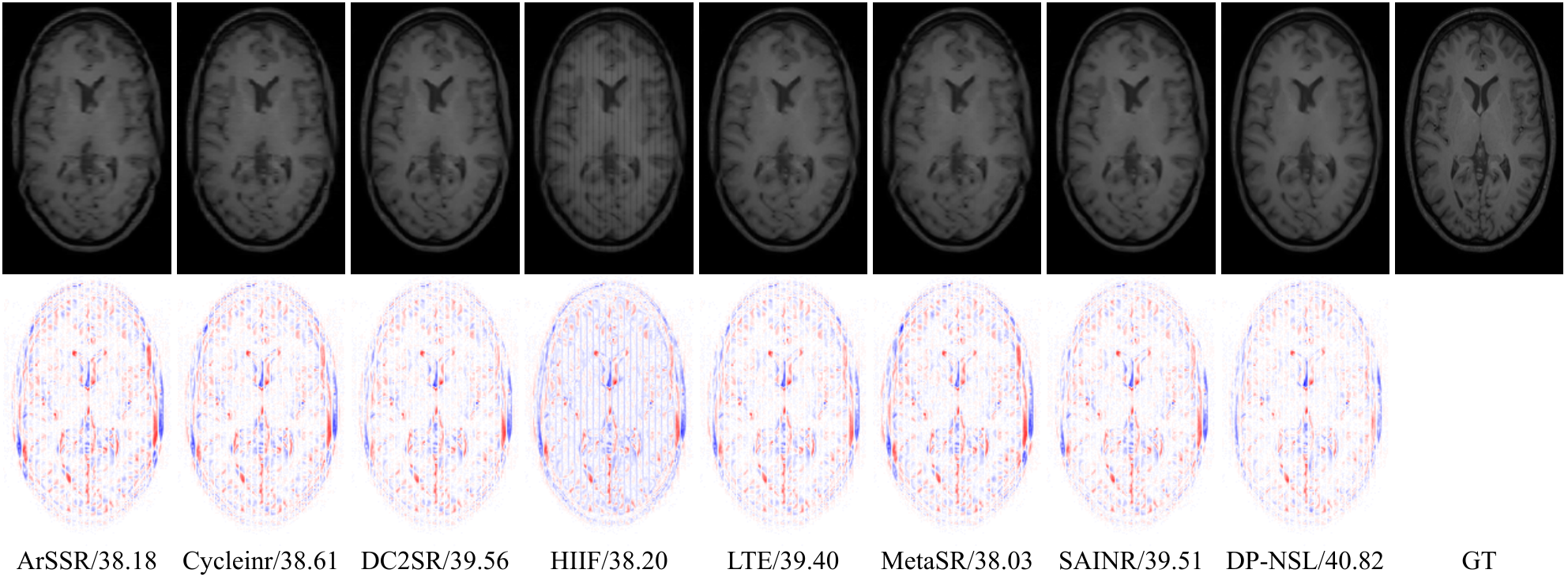}
    \caption{Visualization on the IXI brain MRI dataset (axial view, $\times$5) with error maps.}
    \label{fig:visual_ixi_axis}
\end{figure*}

\noindent\textbf{Qualitative Comparison.}
Fig.~\ref{fig:visual_liver_sag} presents sagittal-view comparisons on the Liver dataset at $\times$3.
It directly exposes through-plane reconstruction quality.
Other methods produce visibly blurred organ boundaries and fail to separate intervertebral spaces clearly. 
By contrast, the proposed DP-NSL recovers smooth, anatomically coherent contours.
Fig.~\ref{fig:visual_ixi_axis} shows axial-view comparisons on the IXI brain MRI dataset at $\times$5.
Cortical folding patterns and white-matter boundaries pose particular challenges due to their convoluted geometry. While other methods blur sulcal regions, DP-NSL reconstructs these complex structures with less error, demonstrating that the locally adaptive geometric prior generalizes to highly textured brain MRI.

\noindent\textbf{Downstream Tasks.}
Table~\ref{tab:seg_kits19} reports downstream segmentation results on KiTS19~\cite{heller2019kits19}. 
DP-NSL achieves the highest Dice score at both $\times$3 and $\times$6, indicating that it preserves the anatomical details for downstream analysis.

\begin{table}[t]

\centering
\caption{Downstream segmentation results (Dice / PSNR) on KiTS19.}
\label{tab:seg_kits19}

\setlength{\tabcolsep}{2pt}

\resizebox{1\linewidth}{!}{
\begin{tabular}{l|cccccccc}
\hline
Scale & MetaSR & LTE & HIIF & ArSSR & SAINR & CycleINR & DC2SR & DP-NSL \\
\hline
$\times$3 & 0.8562/39.77 & 0.8566/40.03 & 0.8570/40.31 & 0.8549/39.95 & 0.8569/39.68 & 0.8561/40.15 & 0.8569/40.04 & 0.8581/40.28 \\
\hline
$\times$6 & 0.7820/34.88 & 0.7891/35.08 & 0.7044/31.50 & 0.7634/34.86 & 0.7922/35.08 & 0.7820/34.98 & 0.7794/35.05 & 0.8067/35.13 \\
\hline
\end{tabular}
}

\end{table}

\subsection{Ablation Studies}
\label{sec:ablation}

Ablation experiments validate the contribution of each proposed component.

\noindent\textbf{Component Analysis.}
Table~\ref{tab:ablation_component} examines the contribution of each core module: MCP, MoS, and LSCD, with PSNR and SSIM.
Without the proposed modules, the baseline degenerates to direct estimation with trilinear upsampling and EDSR, performing unconstrained synthesis.
Activating MCP alone improves PSNR by 0.31~dB at $\times$2, showing that a measurement consistency anchor already provides a noticeable boost without architectural changes.
Replacing the trilinear upsampler with MoS evaluates the macroscopic geometric prior, improving PSNR by 0.63~dB at $\times$2.
Combining MoS with MCP can further improve performance. This supports the hypothesis that, while MoS provides strong detail generation, the null-space projection confines the generated details and prevents them from overriding known geometry.
Finally, the LSCD adds the local inductive bias. The complete DP-NSL achieves the best performance by satisfying both data fidelity and spatial continuity constraints.

\begin{table}[t]
\centering
\caption{Component ablation on the Colon dataset with PSNR and SSIM.}
\label{tab:ablation_component}
\newcommand{\resstd}[2]{#1$\pm$#2}

\begin{tabular}{ccc|cccc}
    \hline
    {MCP} & {MoS} & {LSCD} & $\times 2$ (mean$\pm$std) & $\times 3$ (mean$\pm$std) & $\times 4$ (mean$\pm$std) & $\times 5$ (mean$\pm$std) \\
    \hline
    \ding{55} & \ding{55} & \ding{55} & \resstd{41.82}{0.02} & \resstd{37.76}{0.03} & \resstd{35.30}{0.02} & \resstd{33.64}{0.01} \\
    \ding{51} & \ding{55} & \ding{55} & \resstd{42.13}{0.06} & \resstd{38.03}{0.06} & \resstd{35.52}{0.06} & \resstd{33.80}{0.04} \\
    \ding{55} & \ding{51} & \ding{55} & \resstd{42.45}{0.02} & \resstd{38.32}{0.03} & \resstd{35.82}{0.04} & \resstd{33.96}{0.08} \\
    \ding{55} & \ding{55} & \ding{51} & \resstd{42.06}{0.01} & \resstd{37.95}{0.02} & \resstd{35.43}{0.01} & \resstd{33.67}{0.02} \\
    \ding{51} & \ding{51} & \ding{55} & \resstd{42.52}{0.01} & \resstd{38.40}{0.01} & \resstd{35.89}{0.01} & \resstd{34.08}{0.02} \\
    \ding{51} & \ding{51} & \ding{51} & \resstd{42.54}{0.03} & \resstd{38.42}{0.02} & \resstd{35.93}{0.03} & \resstd{34.12}{0.03} \\
    \hline
    \ding{55} & \ding{55} & \ding{55} & \resstd{0.9810}{0.0001} & \resstd{0.9605}{0.0002} & \resstd{0.9412}{0.0002} & \resstd{0.9232}{0.0004} \\
    \ding{51} & \ding{55} & \ding{55} & \resstd{0.9821}{0.0002} & \resstd{0.9628}{0.0003} & \resstd{0.9440}{0.0004} & \resstd{0.9261}{0.0004} \\
    \ding{55} & \ding{51} & \ding{55} & \resstd{0.9829}{0.0001} & \resstd{0.9641}{0.0001} & \resstd{0.9461}{0.0003} & \resstd{0.9269}{0.0010} \\
    \ding{55} & \ding{55} & \ding{51} & \resstd{0.9817}{0.0001} & \resstd{0.9618}{0.0003} & \resstd{0.9420}{0.0003} & \resstd{0.9227}{0.0004} \\
    \ding{51} & \ding{51} & \ding{55} & \resstd{0.9831}{0.0001} & \resstd{0.9648}{0.0002} & \resstd{0.9470}{0.0003} & \resstd{0.9290}{0.0004} \\
    \ding{51} & \ding{51} & \ding{51} & \resstd{0.9832}{0.0001} & \resstd{0.9648}{0.0002} & \resstd{0.9470}{0.0003} & \resstd{0.9290}{0.0003} \\
    \hline
\end{tabular}

\end{table}

\noindent\textbf{Efficacy of the Multi-Order Geometric Prior.}
Table~\ref{tab:ablation_bspline} investigates the impact of B-spline order within the MoS.
Single-order configurations perform comparably, yielding incremental improvements in reconstruction fidelity from order~1 to~3; however, order~4 offers no additional in-scales benefit and slightly degrades the out-of-scales results.
Multi-order MoS variants that combine spline experts of different orders consistently outperform the single-order variants, supporting the premise that anatomical regions with heterogeneous smoothness require spatially varying geometric priors.
The combination of orders $p \in \{2,3,4\}$ achieves the best in-scales accuracy, while including order~4 is marginally superior at $\times$5.
The tri-order ensemble is therefore adopted as the default configuration, balancing boundary preservation with high-order smoothness.

\begin{figure}[t]
    \centering
    \includegraphics[width=1\textwidth]{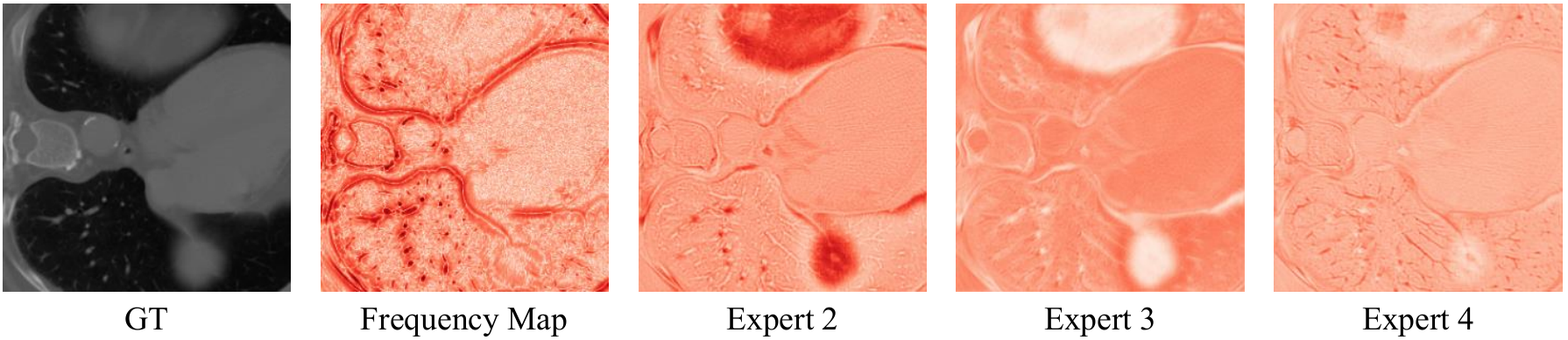}
    \caption{Visualization of MoS routing weights on the CT slice. Frequency maps show the spatial variation intensity of the slice. Expert~$n$ shows spatial activation of the B-spline Upsampler with order $p=n$.}
    \label{fig:routing}
\end{figure}

\begin{table}[t]
\centering
\caption{Ablation study on B-spline order combinations in the MoS module. $p=n$ denotes the expert of order $n$.}
\label{tab:ablation_bspline}
\begin{tabular}{c|cccc}
\hline
\multirow{2}{*}{Orders of MoS} & \multicolumn{4}{c}{PSNR} \\
& $\times$2 & $\times$3 & $\times$4 & $\times$5 \\
\hline
$p=1$ & 42.16 & 38.09 & 35.59 & 33.80 \\
$p=2$ & 42.22 & 38.12 & 35.62 & 33.82 \\
$p=3$ & 42.23 & 38.14 & 35.65 & 33.81 \\
$p=4$ & 42.23 & 38.14 & 35.64 & 33.73 \\
$p \in \{2,3,4\}$ & 42.39 & \textbf{38.30} & \textbf{35.80} & 33.94 \\
$p \in \{1,2,3\}$ & \textbf{42.40} & {38.29} & {35.79} & 33.91 \\
$p \in \{1,2,3,4\}$ & 42.32 & 38.23 & 35.75 & \textbf{33.97} \\
\hline
\end{tabular}
\end{table}

\noindent\textbf{Interplay of Range and Null-Space Operators.}
Table~\ref{tab:ablation_nullspace} dissects the sampling operator choices across three progressive paradigms evaluated via the Range Anchor $\mathcal{U}$ and MCP projection $\Pi_{\mathcal{N}}$. The Range Anchor $\mathcal{U}$ provides the deterministic low-frequency scaffold, while the MCP projection $\Pi_{\mathcal{N}}$ bounds the high-frequency synthesis within the null space.
Without both components, the framework reduces to a pure black-box estimator that must synthesize the low-frequency topology from scratch, undermining stability.
Introducing only a trilinear range anchor $\mathcal{U}$ converts the model to a residual-based architecture. This secures a reliable low-frequency scaffold, yet performance remains sub-optimal due to unconstrained high-frequency reconstruction.
Adding the MCP projection $\Pi_{\mathcal{N}}$ completes the NRSD framework. Within this complete formulation, pairing a trilinear anchor with a zero-padding projection proves optimal. This confirms that well-paired orthogonal operators are needed to balance measurement fidelity with null-space expressiveness.

\subsection{Analysis and Discussion}
\label{sec:analysis}

\noindent\textbf{Expert Routing Interpretation.}
Fig.~\ref{fig:routing} visualizes per-expert routing weight maps on a representative CT slice to investigate whether the MoS router intrinsically learns semantically meaningful allocations.
Two distinct spatial behaviors emerge.
The low-order expert is more active in homogeneous soft-tissue interiors, providing baseline approximations across smooth anatomical regions.
Higher-order experts concentrate on complex structural sites such as fine vascular branches and sharp organ boundaries, capturing high-frequency structural details.
This content-aware routing emerges from data-driven optimization alone, without explicit boundary supervision, indicating that the optimal geometric prior varies by region.
Transitional zones show more balanced expert mixtures, where the router blends geometric priors to match the local structural complexity.

\begin{figure}[t]
    \centering
    \begin{minipage}[c]{0.40\linewidth}
        \centering
        \includegraphics[width=\linewidth]{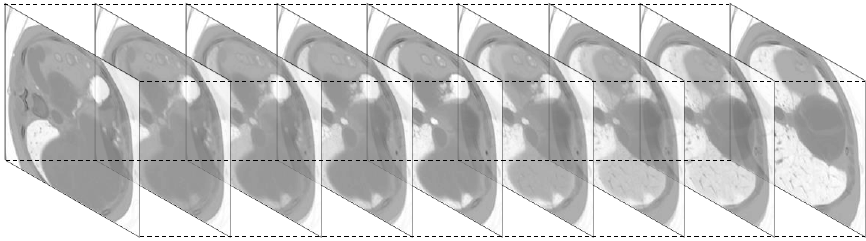}
        \\
        {\footnotesize (a) Range-space Anchor.}

        \vspace{1.6em}
        \includegraphics[width=\linewidth]{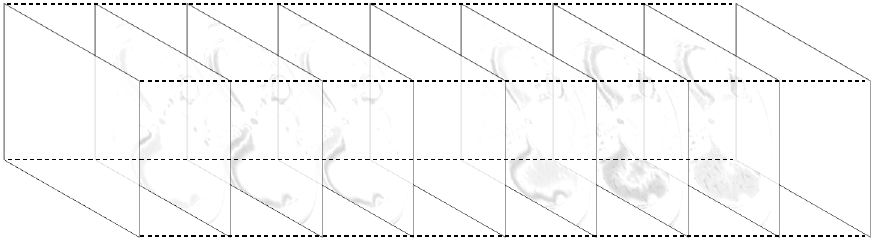}
        \\
        {\footnotesize (b) Null-space component.}
    \end{minipage}
    \hfill
    \begin{minipage}[c]{0.58\linewidth}
        \centering
        \includegraphics[width=\linewidth]{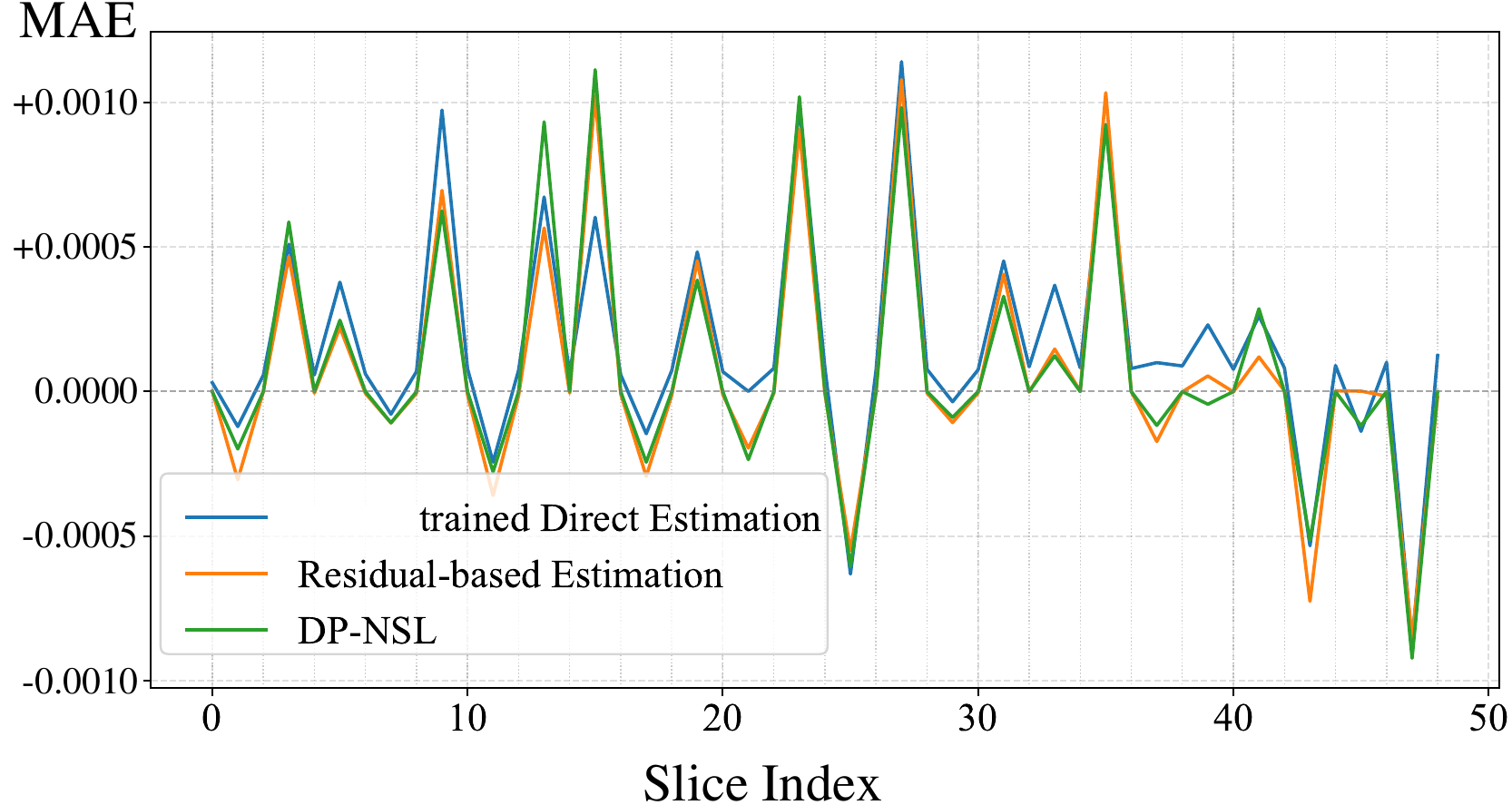}
        \\
        {\footnotesize (c) Slice-wise MAE curve on colon at $\times$4.}
    \end{minipage}
    \caption{Joint visualization of MCP behavior. (a) and (b) show the decomposition into range-space anchor and null-space component respectively. (c) reports slice-wise MAE versus slice index. 
    }
    \label{fig:mcp_consistency}

\end{figure}

\begin{table}[t]
\centering
\caption{Ablation study on the null-range-space operators. ``-'' indicates the corresponding component is omitted.}
\label{tab:ablation_nullspace}
\setlength{\tabcolsep}{2.5mm}
\begin{tabular}{cc|cccc}
\hline
\multirow{2}{*}{Anchor $\mathcal{U}$} & \multirow{2}{*}{Projection $\Pi_{\mathcal{N}}$} & \multicolumn{4}{c}{PSNR} \\
& & $\times$2 & $\times$3 & $\times$4 & $\times$5 \\
\hline
- & - & 41.45 & 37.35 & 34.91 & 33.31 \\
Trilinear & - & 41.76 & 37.68 & 35.21 & 33.54 \\
Zero-Pad & - & 41.11 & 37.08 & 34.69 & 33.56 \\
Zero-Pad & Zero-Pad & 41.89 & 37.78 & 35.28 & 33.63 \\
Trilinear & Trilinear & 41.83 & 36.92 & 34.60 & 32.41 \\
Trilinear & Zero-Pad & \textbf{42.06} & \textbf{37.96} & \textbf{35.45} & \textbf{33.76} \\
\hline
\end{tabular}

\end{table}

\noindent\textbf{Measurement Consistency Verification.}
A central claim of the DP-NSL framework is that MCP guarantees exact data fidelity while confining all learned high-frequency content to the null space.
Fig.~\ref{fig:mcp_consistency} jointly visualizes this decomposition and its quantitative implications.

Fig.~\ref{fig:mcp_consistency}(a) depicts the range-space component $\mathbf{V}_{\mathcal{R}} = \mathcal{U}(\mathbf{V}_{LR})$, the deterministic anchor obtained via trilinear upsampling.
This component reproduces the coarse anatomical layout and faithfully retains all originally observed slice intensities.
Fig.~\ref{fig:mcp_consistency}(b) displays the corresponding null-space component $\mathbf{V}_{\mathcal{N}}$. Safeguarded by MCP, the NSE synthesizes high-frequency anatomical details without generating any superfluous content at the observed slice positions.

Fig.~\ref{fig:mcp_consistency}(c) quantifies this behavior through per-slice Mean Absolute Error (MAE) along the slice axis.
Because the observation constraint is enforced, the MAE drops to zero at the originally acquired positions ($s = 0, R, 2R, \ldots$). This periodic zero-error pattern proves that $\mathcal{D}(\mathbf{V}_{SR}) = \mathbf{V}_{LR}$ holds identically.
By contrast, baseline methods lacking MCP operate as unconstrained direct estimation and residual-based estimation: they exhibit non-negligible errors even at inherently known positions. Such unconstrained behavior subtly alters the original clinical data, presenting an unacceptable risk in diagnostic workflows.

\section{Conclusion}
This paper formulates arbitrary medical slice super-resolution as a constrained inverse problem and presents the DP-NSL framework, which grounds the reconstruction in two complementary priors rather than unconstrained residual-based regression.
The \emph{Deterministic Observation Prior}, enforced by MCP, mathematically guarantees that every originally acquired clinical slice remains unchanged, separating the observable range space from the learnable null space.
The \emph{Geometric Continuity Prior}, instantiated by MoS, regularizes the null-space synthesis with spatially adaptive B-spline bases whose analytic orders adapt to local anatomical complexity. Together with the LSCD for local spatial coherence, these two priors turn an ill-posed reconstruction problem into a well-constrained process reconciling data fidelity with anatomically plausible detail synthesis.
Experiments across three CT and one MRI datasets confirm consistent improvements at both in-scale and out-of-scale factors.
Future work can explore adapting the MCP formulation to more realistic degradation scenarios. 

\section*{Acknowledgements}
This work was supported by the National Natural Science Foundation of China (Grant No. 62471182), Science and Technology Commission of Shanghai Municipality Basic Research Program (Grant No. 25JD1401300), Shanghai Rising-Star Program (Grant No. 24QA2702100), and the Science and Technology Commission of Shanghai Municipality (Grant No. 22DZ2229004)

%
%
\bibliographystyle{splncs04}
\bibliography{main}

@String(CVPR  = {IEEE Conf. Comput. Vis. Pattern Recog.})

@String(NeurIPS = {Adv. Neural Inform. Process. Syst.})

@String(ICML  = {Int. Conf. Mach. Learn.})

@String(ICLR  = {Int. Conf. Learn. Represent.})

@String(CVPRW = {IEEE Conf. Comput. Vis. Pattern Recog. Worksh.})

@String(IJCAI = {IJCAI})

@String(TIP   = {IEEE Trans. Image Process.})

@String(ACMMM = {ACM Int. Conf. Multimedia})

@String(MICCAI = {MICCAI})

@String(TMI   = {IEEE TMI})

@String(IJBHI = {IEEE J. Biomed. Health Inform.})

@String(CVPR  = {CVPR})

@String(NeurIPS = {NeurIPS})

@String(ICML  = {ICML})

@String(ICLR  = {ICLR})

@String(CVPRW = {CVPRW})

@String(TIP   = {IEEE TIP})

@String(ACMMM = {ACM MM})

@String(IJBHI = {IEEE JBHI})

@inproceedings{peng2020saint,
author = {Peng, Cheng and Lin, Wei-An and Liao, Haofu and Chellappa, Rama and Zhou, S Kevin},
title = {Saint: spatially aware interpolation network for medical slice synthesis},
booktitle = CVPR,
year = 2020
}

@inproceedings{lim2017enhanced,
author = {Lim, Bee and Son, Sanghyun and Kim, Heewon and Nah, Seungjun and Mu Lee, Kyoung},
title = {Enhanced deep residual networks for single image super-resolution},
booktitle = CVPRW,
year = 2017
}

@inproceedings{yu2022rplhr,
author = {Yu, Pengxin and Zhang, Haoyue and Kang, Han and Tang, Wen and Arnold, Corey W and Zhang, Rongguo},
title = {RPLHR-CT Dataset and Transformer Baseline for Volumetric Super-Resolution from CT Scans},
booktitle = MICCAI,
year = 2022
}

@inproceedings{hu2019meta,
author = {Hu, Xuecai and Mu, Haoyuan and Zhang, Xiangyu and Wang, Zilei and Tan, Tieniu and Sun, Jian},
title = {Meta-SR: A magnification-arbitrary network for super-resolution},
booktitle = CVPR,
year = 2019
}

@article{antonelli2022medical,
author = {Antonelli, Michela and Reinke, Annika and Bakas, Spyridon and Farahani, Keyvan and Kopp-Schneider, Annette and Landman, Bennett A and Litjens, Geert and Menze, Bjoern and Ronneberger, Olaf and Summers, Ronald M and others},
title = {The medical segmentation decathlon},
journal = {Nature Communications},
volume = 13,
number = 1,
pages = {4128},
year = 2022
}

@article{kingma2014adam,
author = {Kingma, Diederik P and Ba, Jimmy},
title = {Adam: A method for stochastic optimization},
journal = {arXiv preprint arXiv:1412.6980},
year = 2014
}

@article{heller2019kits19,
author = {Heller, Nicholas and Sathianathen, Niranjan and Kalapara, Arveen and Walczak, Edward and Moore, Keenan and Kaluzniak, Heather and Rosenberg, Joel and Blake, Paul and Rengel, Zachary and Oestreich, Makinna and others},
title = {The kits19 challenge data: 300 kidney tumor cases with clinical context, ct semantic segmentations, and surgical outcomes},
journal = {arXiv preprint arXiv:1904.00445},
year = 2019
}

@article{wu2022arbitrary,
author = {Wu, Qing and Li, Yuwei and Sun, Yawen and Zhou, Yan and Wei, Hongjiang and Yu, Jingyi and Zhang, Yuyao},
title = {An arbitrary scale super-resolution approach for 3d mr images via implicit neural representation},
journal = IJBHI,
volume = 27,
number = 2,
pages = {1004--1015},
year = 2022
}

@inproceedings{chen2021learning,
author = {Chen, Yinbo and Liu, Sifei and Wang, Xiaolong},
title = {Learning continuous image representation with local implicit image function},
booktitle = CVPR,
year = 2021
}

@inproceedings{lee2022local,
author = {Lee, Jaewon and Jin, Kyong Hwan},
title = {Local Texture Estimator for Implicit Representation Function},
booktitle = CVPR,
year = 2022
}

@inproceedings{pak2023btc,
author = {Pak, Byeonghyun and Lee, Jaewon and Jin, Kyong Hwan},
title = {B-spline Texture Coefficients Estimator for Screen Content Image Super-Resolution},
booktitle = CVPR,
year = 2023
}

@inproceedings{cao2023ciaosr,
author = {Cao, Jiezhang and Wang, Qin and Xian, Yongqin and Van Gool, Luc and Timofte, Radu},
title = {CiaoSR: Continuous Implicit Attention-in-Attention Network for Arbitrary-Scale Image Super-Resolution},
booktitle = CVPR,
year = 2023
}

@inproceedings{jiang2025hiif,
author = {Jiang, Yuxuan and Kwan, Ho Man and Zhang, Fan and Bull, David},
title = {HIIF: Hierarchical Encoding based Implicit Image Function for Continuous Super-resolution},
booktitle = CVPR,
year = 2025
}

@inproceedings{he2024lmf,
author = {He, Zongyao and Jin, Zhi},
title = {Latent Modulated Function for Computational Optimal Continuous Image Representation},
booktitle = CVPR,
year = 2024
}

@inproceedings{fang2024cycleinr,
author = {Fang, Wei and Tang, Yuxing and Xu, Minfeng and Pan, Jing and Cai, Jianping},
title = {CycleINR: Cycle Implicit Neural Representation for Arbitrary-Scale Volumetric Super-Resolution of Medical Data},
booktitle = CVPR,
year = 2024
}

@inproceedings{zeng2025dc2sr,
author = {Zeng, Chuan and Zhang, Zhao and Huang, Wei and Zhang, Lei},
title = {DC2-SR: A Dual-Consistency Guided Curriculum Learning Method for Thick-Slice Fetal MRI Super-Resolution},
booktitle = ACMMM,
year = 2025
}

@article{song2024i3net,
author = {Song, Haofei and Mao, Xintian and Wang, Yan and Shen, Wei and Li, Qingli},
title = {I3Net: Inter-Intra-slice Interpolation Network for Medical Slice Synthesis},
journal = TMI,
year = 2024
}

@article{wang2023sainr,
title={Spatial attention-based implicit neural representation for arbitrary reduction of MRI slice spacing},
author={Wang, Xin and Wang, Sheng and Xiong, Honglin and Xuan, Kai and Zhuang, Zixu and Liu, Mengjun and Shen, Zhenrong and Zhao, Xiangyu and Zhang, Lichi and Wang, Qian},
journal={Medical Image Analysis},
volume={94},
pages={103158},
year={2024},
publisher={Elsevier}
}

@inproceedings{kim2025bfstvsr,
author = {Kim, Eunjin and Kim, Hyeonjin and Yoo, Jaejun},
title = {BF-STVSR: B-Splines and Fourier -- Best Friends for High Fidelity Spatial-Temporal Video Super-Resolution},
booktitle = CVPR,
year = 2025
}

@inproceedings{wang2021spe,
author = {Wang, Peng-Shuai and Liu, Yang and Yang, Yu-Qi and Tong, Xin},
title = {Spline Positional Encoding for Learning 3D Implicit Signed Distance Fields},
booktitle = IJCAI,
year = 2021
}

@inproceedings{wang2023ddnm,
author = {Wang, Yinhuai and Yu, Jiwen and Zhang, Jian},
title = {Zero-Shot Image Restoration Using Denoising Diffusion Null-Space Model},
booktitle = ICLR,
year = 2023
}

@inproceedings{jacome2025npn,
author = {Jacome, Roman and Gualdron-Hurtado, Romario and Arguello, Henry},
title = {NPN: Non-Linear Projections of the Null-Space for Imaging Inverse Problems},
booktitle = NeurIPS,
year = 2025
}

@article{chen2025unnull,
author = {Chen, Yurong and Wang, Yaonan and Zhang, Hui},
title = {Unsupervised Range-Nullspace Learning Prior for Multispectral Images Reconstruction},
journal = TIP,
year = 2025
}

@misc{ixi_dataset,
  title={IXI Dataset},
  howpublished={\url{https://brain-development.org/ixi-dataset/}},
}

@article{gerig1992nonlinear,
  title={Nonlinear anisotropic filtering of MRI data},
  author={Gerig, Guido and Kubler, Olaf and Kikinis, Ron and Jolesz, Ferenc A},
  journal={IEEE Trans. Med. Imaging},
  volume={11},
  number={2},
  pages={221--232},
  year={1992},
  publisher={IEEE}
}

@book{birkfellner2014applied,
  title={Applied Medical Image Processing: A Basic Course},
  author={Birkfellner, Wolfgang},
  year={2014},
  publisher={CRC Press}
}

@inproceedings{szegedy2015going,
author = {Szegedy, Christian and Liu, Wei and Jia, Yangqing and Sermanet, Pierre and Reed, Scott and Anguelov, Dragomir and Erhan, Dumitru and Vanhoucke, Vincent and Rabinovich, Andrew},
title = {Going Deeper with Convolutions},
booktitle = CVPR,
year = 2015
}

@inproceedings{szegedy2016rethinking,
author = {Szegedy, Christian and Vanhoucke, Vincent and Ioffe, Sergey and Shlens, Jonathon and Wojna, Zbigniew},
title = {Rethinking the Inception Architecture for Computer Vision},
booktitle = CVPR,
year = 2016
}

@inproceedings{wei2023srno,
  title={Super-resolution neural operator},
  author={Wei, Min and Zhang, Xuesong},
  booktitle=CVPR,
  pages={18247--18256},
  year={2023}
}

@inproceedings{luo2024hinote,
  title={Hierarchical Neural Operator Transformer with Learnable Frequency-aware Loss Prior for Arbitrary-scale Super-resolution},
  author={Luo, Xihaier and Qian, Xiaoning and Yoon, Byung-Jun},
  booktitle=ICML,
  year={2024}
}

@inproceedings{liu2025difffno,
  title={Difffno: Diffusion fourier neural operator},
  author={Liu, Xiaoyi and Tang, Hao},
  booktitle=CVPR,
  pages={150--160},
  year={2025}
}

@article{uhm2025acvtt,
  title={An Anisotropic Cross-View Texture Transfer with Multi-Reference Non-Local Attention for CT Slice Interpolation},
  author={Uhm, Kwang-Hyun and Cho, Hyunjun and Hong, Sung-Hoo and Jung, Seung-Won},
  journal={IEEE Trans. Med. Imaging},
  year={2025},
  publisher={IEEE}
}

@article{unser2002bspline_theory_1,
  title={B-spline signal processing. I. Theory},
  author={Unser, Michael and Aldroubi, Akram and Eden, Murray},
  journal={IEEE transactions on signal processing},
  volume={41},
  number={2},
  pages={821--833},
  year={2002},
  publisher={IEEE}
}

@article{unser2002bspline_theory_2,
  title={B-spline signal processing. II. Efficiency design and applications},
  author={Unser, Michael and Aldroubi, Akram and Eden, Murray},
  journal={IEEE transactions on signal processing},
  volume={41},
  number={2},
  pages={834--848},
  year={2002},
  publisher={IEEE}
}

\clearpage 
\appendix  

\counterwithin{equation}{section}
\counterwithin{figure}{section}
\counterwithin{table}{section}

\begin{center}
    \Large \bf Supplementary Material for Dual-Prior Guided Null-Space Learning with Mixture-of-Splines for Arbitrary Medical Slice Super-Resolution
\end{center}

This supplementary material provides additional implementation details and experiments. 
First, Sec.~\ref{sec:supp_bspline} formalizes the B-spline basis functions used in the Mixture-of-Splines (MoS) module. 
Sec.~\ref{sec:supp_fixscale} reports fixed-scale training results to isolate each method's fitting capacity at specific resolutions.
Sec.~\ref{sec:supp_lscd} compares the Local Spatial Consistency Decoder (LSCD) against alternative architectures. 
Sec.~\ref{sec:supp_cost} breaks down computational costs, Sec.~\ref{sec:supp_epoch3000} examines longer training, and Sec.~\ref{sec:supp_vis} includes more visual comparisons.

\section{B-Spline Basis Function Definitions}
\label{sec:supp_bspline}

The Mixture-of-Splines (MoS) in DP-NSL employs B-spline basis functions of different orders as geometric priors.
A B-spline basis function of order $p$ guarantees $C^{p-1}$ analytic continuity and has compact support over the interval $[-(p+1)/2,\;(p+1)/2]$.
Formally, the B-spline basis is defined recursively through repeated convolution of the zeroth-order box function~\cite{unser2002bspline_theory_1,unser2002bspline_theory_2}:
\begin{equation}
    \beta^{0}(x) = 
    \begin{cases}
        1, & |x| < \tfrac{1}{2}, \\
        0, & \text{otherwise},
    \end{cases}
    \qquad
    \beta^{p}(x) = \underbrace{\beta^{0} * \beta^{0} * \cdots * \beta^{0}}_{(p+1)\text{ times}}(x).
\end{equation}
This recursive convolution yields piecewise polynomial functions with increasing smoothness.
The explicit piecewise forms used in MoS (orders $p \in \{2,3,4\}$) are given below.


\noindent\textbf{Second-order basis} (supported in $[-\tfrac{3}{2},\,\tfrac{3}{2}]$):
\begin{equation}
    \beta^{2}(x) = 
    \begin{cases}
        \tfrac{1}{2}\big(\tfrac{3}{2} + x\big)^{2}, & -\tfrac{3}{2} \leq x \leq -\tfrac{1}{2}, \\[4pt]
        \tfrac{3}{4} - x^{2}, & -\tfrac{1}{2} < x \leq \tfrac{1}{2}, \\[4pt]
        \tfrac{1}{2}\big(\tfrac{3}{2} - x\big)^{2}, & \tfrac{1}{2} < x \leq \tfrac{3}{2}, \\[4pt]
        0, & \text{otherwise}.
    \end{cases}
\end{equation}

\noindent\textbf{Third-order basis} (supported in $[-2,\,2]$):
\begin{equation}
    \beta^{3}(x) = 
    \begin{cases}
        \tfrac{1}{6}(2 + x)^{3}, & -2 \leq x \leq -1, \\[4pt]
        \tfrac{1}{6}\big(4 - 6x^{2} - 3x^{3}\big), & -1 < x \leq 0, \\[4pt]
        \tfrac{1}{6}\big(4 - 6x^{2} + 3x^{3}\big), & 0 < x \leq 1, \\[4pt]
        \tfrac{1}{6}(2 - x)^{3}, & 1 < x \leq 2, \\[4pt]
        0, & \text{otherwise}.
    \end{cases}
\end{equation}

\noindent\textbf{Fourth-order basis} (supported in $[-\tfrac{5}{2},\,\tfrac{5}{2}]$):
\begin{equation}
    \beta^{4}(x) = \frac{1}{24}
    \begin{cases}
        \big(\tfrac{5}{2} + x\big)^{4}, & -\tfrac{5}{2} \leq x \leq -\tfrac{3}{2}, \\[4pt]
        -4x^{4} - 20x^{3} - 30x^{2} - 5x + \tfrac{55}{4}, & -\tfrac{3}{2} < x \leq -\tfrac{1}{2}, \\[4pt]
        6x^{4} - 15x^{2} + \tfrac{115}{8}, & -\tfrac{1}{2} < x \leq \tfrac{1}{2}, \\[4pt]
        -4x^{4} + 20x^{3} - 30x^{2} + 5x + \tfrac{55}{4}, & \tfrac{1}{2} < x \leq \tfrac{3}{2}, \\[4pt]
        \big(\tfrac{5}{2} - x\big)^{4}, & \tfrac{3}{2} < x \leq \tfrac{5}{2}, \\[4pt]
        0, & \text{otherwise}.
    \end{cases}
\end{equation}

As described in the main paper, each B-spline upsampling expert $\mathcal{U}_{p}$ in MoS evaluates the corresponding basis $\beta^{p}$.

\section{Comparative Results at Fixed Upsampling Factors}
\label{sec:supp_fixscale}

We train and evaluate each method independently at single fixed upsampling factors. This isolates each model's per-scale fitting capacity from the broader challenge of arbitrary-scale generalization. Table~\ref{tab:fixed_scale} reports the results. DP-NSL achieves the highest PSNR and SSIM across all four scales.

\begin{table}[!h]
\centering
\caption{Fixed-scale quantitative comparison. Best results in \textbf{bold}, second best \underline{underlined}.}
\label{tab:fixed_scale}

\renewcommand{\arraystretch}{1}

\resizebox{1\linewidth}{!}{
\begin{tabular}{l|c|c|c|c}
\toprule
\multirow{2}{*}{Method} & $\times2$ & $\times3$ & $\times4$ & $\times5$ \\
 & PSNR/SSIM & PSNR/SSIM & PSNR/SSIM & PSNR/SSIM \\
\midrule
EDSR3D   & {41.84}/0.9809 & 37.71/0.9600 & {35.37}/{0.9421} & 33.99/0.9281 \\
MetaSR   & 41.31/0.9790 & 36.98/0.9549 & 35.16/0.9397 & 33.56/0.9227 \\
LTE      & 42.12/0.9818 & 37.92/\underline{0.9615} & 35.56/0.9437 & 34.03/0.9286 \\
HIIF     & 42.00/0.9815 & 37.89/0.9612 & 35.46/0.9429 & 32.40/0.8747 \\
ArSSR    & 40.96/0.9776 & 36.93/0.9545 & 34.65/0.9347 & 33.32/0.9202 \\
SAINR    & 42.05/0.9814 & \underline{37.96}/0.9613 & \underline{35.66}/\underline{0.9443} & \underline{34.14}/\underline{0.9299} \\
CycleINR & 42.03/0.9815 & 37.69/0.9598 & 35.22/0.9403 & 33.73/0.9242 \\
DC2SR    & \underline{42.17}/\underline{0.9820} & 37.82/0.9609 & 35.44/0.9425 & 33.89/0.9271 \\
\textbf{DP-NSL} & \textbf{42.96/0.9842} & \textbf{38.55/0.9656} & \textbf{36.01/0.9480} & \textbf{34.50/0.9340} \\
\bottomrule
\end{tabular}
}
\end{table}


\section{Analysis of LSCD}
\label{sec:supp_lscd}

To justify the Local Spatial Consistency Decoder (LSCD), we test different decoder variants:
\begin{itemize}
    \item \textbf{Pixel-wise Decoder (PWD)}: A standard LIIF-based~\cite{chen2021learning} MLPs that decode features independently at each coordinate.
    \item \textbf{Slice-wise Decoder (SWD)}: A standard convolutional decoder that processes the entire 2D slice.
    \item \textbf{LSCD}: Our multi-scale, inception-style decoder.
\end{itemize}

\begin{table}[!h]
\centering
\caption{Analysis of LSCD, including PSNR (dB), FLOPs (M) and parameters (K).}
\label{tab:ablation_decoder}

\renewcommand{\arraystretch}{1}

\setlength{\tabcolsep}{4pt}

\resizebox{1\linewidth}{!}{

\begin{tabular}{l| r@{/}r@{/}r | r@{/}r@{/}r | r@{/}r@{/}r}
\toprule
\multirow{2}{*}{Method} & \multicolumn{3}{c|}{$\times2$} & \multicolumn{3}{c|}{$\times3$} & \multicolumn{3}{c}{$\times4$} \\
\cmidrule(lr){2-4} \cmidrule(lr){5-7} \cmidrule(lr){8-10}
    & PSNR & FLOPs & Params & PSNR & FLOPs & Params & PSNR & FLOPs & Params \\
\midrule
PWD$_{w\slash o \text{Coord}} $      & 42.16 & 98 & 214 & 38.10 & 140 & 214 & 34.95 & 182 & 214  \\
PWD$_{w\text{Coord}} $  & 42.11 & 98 & 215 & 38.11 & 140 & 215 & 35.53 & 182 & 215  \\
SWD$_{w\slash o \text{Coord}} $    & 41.79 & 52 & 112 & 37.72 & 74 & 112 & 35.28 & 96 & 112  \\
SWD$_{w\text{Coord}} $ & 41.86 & 54 & 118 & 37.79 & 77 & 118 & 35.33 & 100 & 118  \\
LSCD       & 42.05 & 32 & 71 & 37.94 & 46 & 71 & 35.43 & 60 & 71  \\
\bottomrule
\end{tabular}


}
\end{table}

Table~\ref{tab:ablation_decoder} compares their performance. ``$w \text{Coord}$'' and ``$w\slash o \text{Coord}$'' denote whether explicit coordinate grids are appended as position constraints.

Because PWD processes coordinates in isolation, it struggles heavily without explicit coordinate constraints, and stacking MLP layers quickly drives up computation. SWD relies on 2D convolutions to supply spatial structure, making coordinates less critical, but its overall reconstruction quality falls short. By using multi-scale decoding in an inception-style layout, LSCD outperforms both MLP and standard convolution variants while keeping low costs.

\section{Computational Costs}
\label{sec:supp_cost}

Table~\ref{tab:result_cost} compares model complexity and inference times, measured on a single NVIDIA RTX 3090 GPU.

Simple implicit neural representation models like ArSSR and DC$^2$SR run fast but produce smoother, less detailed outputs. On the other hand, SAINR and HIIF rely on complex spatial attention or deep coordinate-wise MLPs. Their FLOPs and latency scale poorly at larger upsampling factors. 
DP-NSL is not the absolute cheapest model, but achieves a favorable balance of high reconstruction performance and reasonable computational costs.

\begin{table}[!h]
\centering
\caption{Computational cost comparison, including FLOPs (G), parameters (K) and inference time (ms).}
\label{tab:result_cost}

\renewcommand{\arraystretch}{1}

\setlength{\tabcolsep}{4pt}

\resizebox{1\linewidth}{!}{
\begin{tabular}{l | r@{/}r@{/}r | r@{/}r@{/}r | r@{/}r@{/}r }
\toprule
Method & \multicolumn{3}{c|}{$\times2$} & \multicolumn{3}{c|}{$\times3$} & \multicolumn{3}{c}{$\times4$} \\
\cmidrule(lr){2-4} \cmidrule(lr){5-7} \cmidrule(lr){8-10}
 & FLOPs & Params & Time & FLOPs & Params & Time & FLOPs & Params & Time \\
\midrule
EDSR3D    & 958  & 3660 & 43 & 959  & 3660 & 43 & 959  & 3660 & 43 \\
MetaSR    & 1160 & 4100 & 227 & 1250 & 4100 & 308 & 1330 & 4100 & 382 \\
LTE       & 1100 & 4080 & 61 & 1130 & 4080 & 68 & 1160 & 4080 & 74 \\
HIIF      & 1710 & 5490 & 335 & 1990 & 5490 & 470 & 2260 & 5490 & 599 \\
ArSSR     & 1060 & 3870 & 51 & 1100 & 3870 & 55 & 1140 & 3870 & 59 \\
SAINR     & 1330 & 3880 & 406 & 1490 & 3880 & 570 & 1650 & 3880 & 716 \\
CycleINR  & 1160 & 3880 & 96 & 1250 & 3880 & 120 & 1330 & 3880 & 143 \\
DC2SR     & 1110 & 4090 & 55 & 1160 & 4090 & 59 & 1200 & 4090 & 63 \\
DP-NSL    & 1250 & 4380 & 98 & 1360 & 4380 & 122 & 1470 & 4380 & 145 \\
\bottomrule
\end{tabular}
}
\end{table}

\section{Convergence Analysis}
\label{sec:supp_epoch3000}

To validate model convergence, ArSSR, CycleINR, and DP-NSL are trained from scratch for 3000 epochs under the same protocol. Table~\ref{tab:E3000} shows that all methods gain only marginally. These limited gains suggest that 1000-epoch training is already close to convergence.

\begin{table}[ht]
\centering
\caption{PSNR and SSIM after 3000 training epochs. Values in parentheses denote gains over the 1000-epoch setting.}
\label{tab:E3000}
\setlength{\tabcolsep}{2.5pt}
\newcommand{\resdelta}[2]{#1{\scriptsize\,(\,#2\,)}}
\resizebox{1\linewidth}{!}{
\begin{tabular}{llcccccc}
\toprule
Metric & Method & $\times2$ & $\times3$ & $\times4$ & $\times5$ & $\times6$ & $\times7$ \\
\midrule
\multirow{3}{*}{PSNR}
& ArSSR & \resdelta{40.75}{+0.08} & \resdelta{36.86}{+0.07} & \resdelta{34.30}{+0.06} & \resdelta{32.81}{+0.03} & \resdelta{31.69}{+0.03} & \resdelta{30.75}{+0.01} \\
& CycleINR & \resdelta{41.49}{+0.14} & \resdelta{37.57}{+0.19} & \resdelta{35.02}{+0.15} & \resdelta{33.07}{+0.19} & \resdelta{31.93}{+0.13} & \resdelta{30.97}{+0.10} \\
& DP-NSL & \resdelta{\textbf{42.62}}{+0.07} & \resdelta{\textbf{38.49}}{+0.06} & \resdelta{\textbf{35.98}}{+0.03} & \resdelta{\textbf{34.17}}{+0.03} & \resdelta{\textbf{32.88}}{+0.02} & \resdelta{\textbf{31.83}}{+0.00} \\
\midrule
\multirow{3}{*}{SSIM}
& ArSSR & \resdelta{0.9774}{+0.0003} & \resdelta{0.9540}{+0.0006} & \resdelta{0.9302}{+0.0009} & \resdelta{0.9120}{+0.0006} & \resdelta{0.8964}{+0.0005} & \resdelta{0.8833}{+0.0003} \\
& CycleINR & \resdelta{0.9800}{+0.0006} & \resdelta{0.9587}{+0.0013} & \resdelta{0.9375}{+0.0015} & \resdelta{0.9144}{+0.0025} & \resdelta{0.8992}{+0.0020} & \resdelta{0.8864}{+0.0018} \\
& DP-NSL & \resdelta{\textbf{0.9834}}{+0.0002} & \resdelta{\textbf{0.9652}}{+0.0004} & \resdelta{\textbf{0.9476}}{+0.0005} & \resdelta{\textbf{0.9297}}{+0.0005} & \resdelta{\textbf{0.9144}}{+0.0005} & \resdelta{\textbf{0.9002}}{+0.0003} \\
\bottomrule
\end{tabular}
}
\end{table}

\section{More Visualizations}
\label{sec:supp_vis}

\begin{figure}[!h]
    \centering
    \includegraphics[width=1\textwidth]{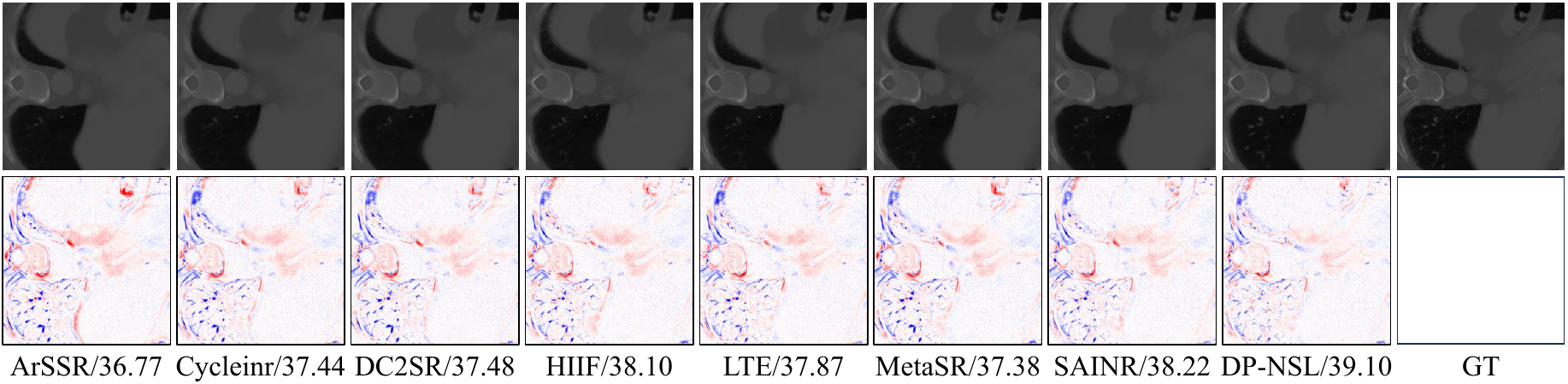}
    \caption{Axial-view visual comparisons on the Colon dataset at $\times$2.}
\end{figure}

\begin{figure}[!h]
    \centering
    \includegraphics[width=1\textwidth]{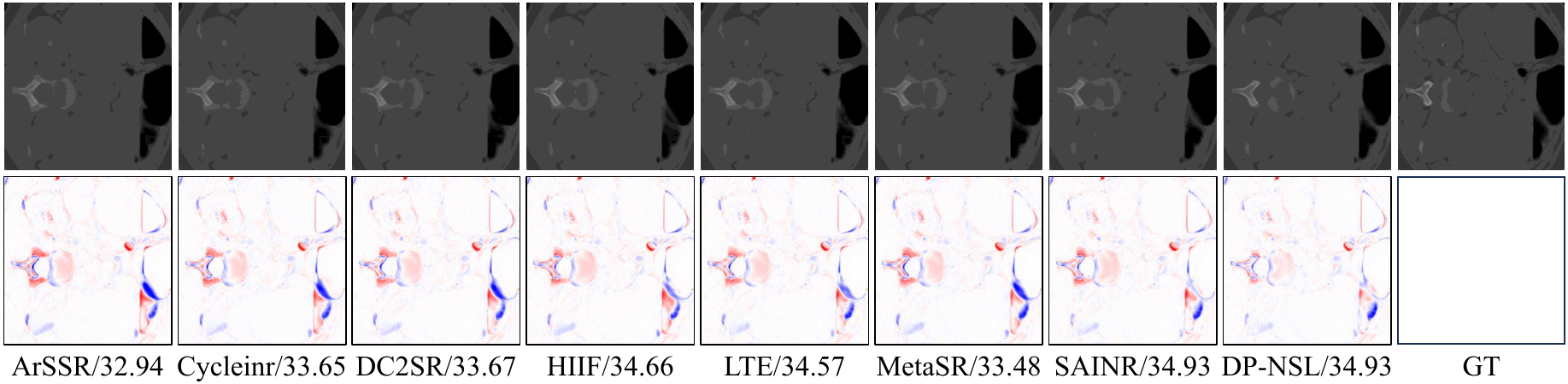}
    \caption{Axial-view visual comparisons on the Colon dataset at $\times$3.}
\end{figure}

\begin{figure}[!h]
    \centering
    \includegraphics[width=1\textwidth]{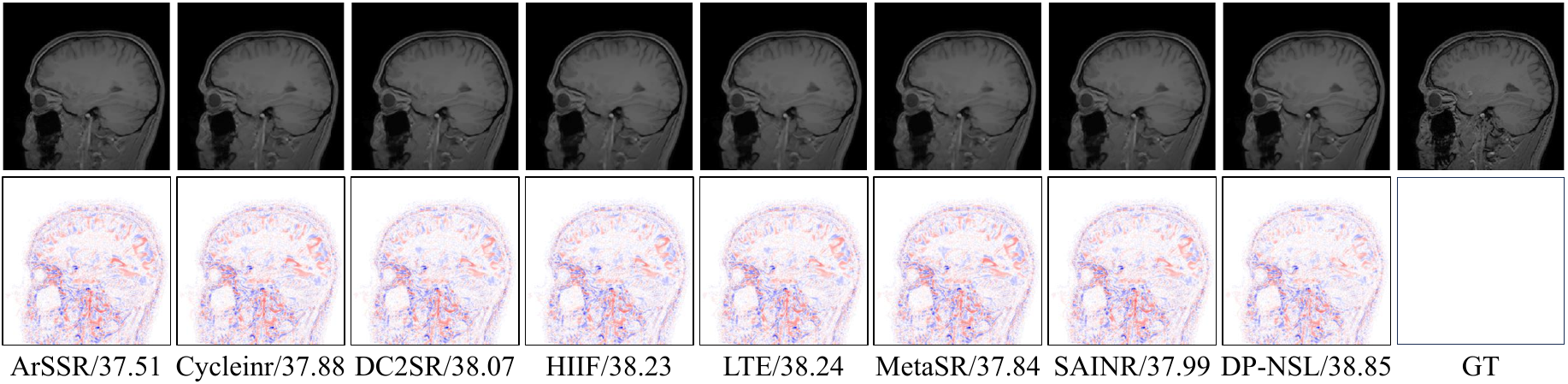}
    \caption{Sagittal-view visual comparisons on the IXI dataset at $\times$4.}
\end{figure}


\end{document}